\documentstyle[epsf]{cupconf}

\title[Population Synthesis at low and high $z$]{Stellar population synthesis models at low and high redshift$^\dagger$}

\author{Gustavo Bruzual A.}

\affiliation{Centro de Investigaciones de Astronom{\'\i}a (CIDA), A.P. 264, M\'erida, Venezuela}

\begin{document}
\ifnfssone
\else
  \ifnfsstwo
  \else
    \ifoldfss
      \let\mathcal\cal
      \let\mathrm\rm
      \let\mathsf\sf
    \fi
  \fi
\fi

\maketitle

\begin{abstract}
The basic assumptions behind Population Synthesis and Spectral Evolution models
are reviewed. The numerical problems encountered by the standard population
synthesis technique when applied to models with truncated star formation rates
are described. The Isochrone Synthesis algorithm is introduced as a means to
circumvent these problems. A summary of results from the application of this
algorithm to model galaxy spectra by Bruzual and Charlot (1993, 2000) follows.
I present a comparison of these population synthesis model predictions
with observed spectra and color magnitude diagrams for stellar systems of
various ages and metallicities. It is argued that models built using different
ingredients differ in the resulting values of some basic quantities (e.g.
$M/L_V$), without need to invoking violations of physical principles. The range
of allowed colors in the observer frame is explored for several galaxy redshifts.
\footnote{To appear in {\it Proceedings of the XI Canary Islands Winter School
of Astrophysics on Galaxies at High Redshift}, eds. I. P\'erez-Fournon,
M. Balcells and F. S\'anchez}
\end{abstract}

\section{Introduction}
The number distribution of the stellar populations present in a galaxy is a
function of time. Thus, the number of stars of a given spectral type, luminosity
class, and metallicity content changes as the galaxy ages.
In early-type galaxies (E/S0) most of the stars were formed during, or very
early after the initial collapse of the galaxy and the stellar population ages
as times goes by. The chemical abundance in these systems must have reached
the value measured in the stars very quickly during the formation process since
most E/S0 galaxies show little evidence of recent major events of star
formation.
In late-type galaxies the stellar population also ages, but there is a
significant number of new stars being formed. Depending on the star formation
rate, $\Psi(t)$, the mean age of the stars in a galaxy may even decrease as
the galaxy gets older. In general, in late-type systems the metal content of
the stars and the interstellar medium is an increasing function of time.
$\Psi(t)$ can also increase above its typical value due to interactions
between two or more galaxies or with the environment.

As a consequence of the aging of the stellar population, or its renewal in
the case of galaxies with high recent $\Psi(t)$,
the rest-frame spectral distribution
of the light emitted by a galaxy is a function of its proper time.
Observational properties such as photometric magnitude and colors, line strength
indices, metal content of gas and young stars, depend on the epoch at which we
observe a galaxy on its reference frame. This {\it intrinsic} evolution should
not be mistaken with the {\it apparent} evolution produced by the cosmological
redshift $z$. For distant galaxies both effects may be equally significant.

In order to search for spectral evolution in galaxy samples we must {\it (a)}
quantify the amount of evolution expected between cosmological epoch $t_1$ and
$t_2$, and {\it (b)} design observational tests that will reveal this amount of
evolution, if present. Evolutionary population synthesis models predict the
amount of evolution expected under different scenarios and allow us to judge
the feasibility of measuring it.
In the ideal case (Arag\'on-Salamanca et al. 1993; Stanford et al. 1995, 1998;
Bender et al. 1996) spectral evolution is measured simply by comparing
spectra of galaxies obtained in such a way that the same rest frame wavelength
region is sampled in all galaxies, irrespective of $z$.
In this case there is no need to apply uncertain $K$-corrections to transform
all the spectra to a common wavelength scale.
Alternatively, if this approach cannot be applied, e.g. when studying large
samples of faint galaxies, one can use models which allow for different degrees
of evolution (including none) to derive indirectly the amount of evolution
consistent with the data (Pozzetti et al. 1996; Metcalfe et al. 1996).
Clearly, the first approach is to be preferred whenever possible.
In all cases we must rule out possible deviations from the natural or passive
evolution of the stellar population in some of the galaxies under study
induced, for example, by interactions with other galaxies or cluster
environment, etc.

The large amount of astrophysical data that has become available in the last few
years has made possible to build several complete sets of stellar population
synthesis models. The predictions of these models have been used to study
many types of stellar systems, from local normal galaxies to the most
distant galaxies discovered so far (approaching $z$ of 4 to 5), from
globular clusters in our galaxy to proto-globular clusters forming in
different environments in distant, interacting galaxies.
In this paper I present an overview of results from population synthesis models
directly applicable to the interpretation of galaxy spectra.

\section{The Population Synthesis Problem}
Stellar evolution theory provides us with the functions $T_{eff}(m,Z,t)$ and
$L(m,Z,t)$ which describe the behavior in time $t$ of the effective temperature
$T_{eff}$ and luminosity $L$ of a star of mass $m$ and metal abundance $Z$.
For fixed $m$ and $Z$, $L(t)$ and $T_{eff}(t)$ describe parametrically in the
H-R diagram the evolutionary track for stars of this mass and metallicity. 
The initial mass function (IMF), $\phi(m)$, indicates the number of stars of
mass $m$
born per unit mass when a stellar population is formed. The star formation rate
(SFR), $\Psi(t)$, gives the amount of mass transformed into stars per unit time
according to $\phi(m)$. The metal enrichment function (MEF), $Z(t)$, also follows
from the theory of stellar evolution.
The population synthesis problem can then be stated as follows.
Given a complete set of evolutionary tracks and the functions
$\phi(m)$ and $\Psi(t)$, compute the number of stars present at each
evolutionary stage in the H-R diagram as a function of time. To solve this
problem exactly we need additional knowledge about the MEF, $Z(t)$, which gives
the time evolution of the chemical abundance
of the gas from which the successive generations of stars are formed.
For simplicity, it is commonly assumed that $\phi(m)$ and $\Psi(t)$ are
decoupled from $Z(t)$, even though it is recognized that in real stellar
systems these three quantities are most likely closely interrelated.
The spectral evolution problem can be solved trivially once the population
synthesis problem is solved, provided that we know the spectral energy
distribution (SED) at each point in the H-R diagram representing an evolutionary
stage in our set of tracks. The discussion that follows is based in the work
of Charlot and Bruzual (1991, hereafter CB91) and Bruzual and Charlot (1993,
2000, hereafter BC93 and BC2000). See also Bruzual (1998, 1999, 2000).
Unless otherwise indicated, I will ignore chemical evolution and assume that
at all epochs stars form with a single metallicity, $Z(t)$ = constant,
and the same $\phi(m)$. 

Let $N_i^o$ be the number of stars of mass $M_i$ born when an instantaneous
burst of star formation occurs at $t=0$. This kind of burst population has
been called simple stellar population (SSP, Renzini 1981). When we look at
this population at later times, we will see the stars traveling along the
corresponding evolutionary track. If the stars live in their $k^{th}$
evolutionary stage from time $t_{i,k-1}$ to time $t_{i,k}$, then at time
$t$ the number of stars of this mass populating the $k^{th}$ stage is simply
$$N_{i,k}(t)=\cases{N_i^o,&if $t_{i,k-1} \le t < t_{i,k}$;\cr
                        0,&otherwise.\cr}\eqno(1)$$
For an arbitrary SFR, $\Psi(t)$, we compute the number of stars $\eta_{i,k}(t)$
of mass $M_i$ at the $k^{th}$ evolutionary stage from the following convolution
integral
$$\eta_{i,k}(t)=\int_0^t\Psi(t-t')N_{i,k}(t')dt',\eqno(2)$$
which in view of (1) can be written as
$$\eta_{i,k}(t)=N_i^o \int_{t_{i,k-1}}^{min(t,t_{i,k})} \Psi(t-t')dt'.\eqno(3)$$
From (3) we see that the commonly heard statement that the number of stars
expected in a stellar population at a given position in the H-R diagram is
proportional to the time spent by the stars at this position, i.e.
$$\eta_{i,k} \propto N_i^o (t_{i,k} - t_{i,k-1}),\eqno(4)$$
is accurate only for a constant $\Psi(t)$.
For non-constant SFRs, the integral in (3) assigns more weight to the epochs
of higher star formation. For instance, for an exponentially decaying SFR with
e-folding time $\tau$, $\Psi(t)=\exp(-t/\tau)$, we have
$$\eta_{i,k}(t)\propto
N_i^o \{\exp[-(t-t_{i,k})/\tau] - \exp[-(t-t_{i,k-1})/\tau]\}.\eqno(5)$$
$\Psi(t)$ was stronger at time $(t-t_{i,k})$ than at time $(t-t_{i,k-1})$,
which is clearly taken into account in (5).

A prerequisite for building trustworthy population synthesis and spectral 
evolution models is an adequate algorithm to follow the evolution of
consecutive generations of stars in the H-R diagram. This goal is
accomplished by the standard technique described above
provided that the function $\Psi(t)$ extends from $t=0$ to $t=\infty$. 
Special caution is required if $\Psi(t)$ becomes $0$ at a finite age.
As an illustration, let us consider the case of a burst of star formation
which lasts for a finite length of time $\tau$, 
$$\Psi(t)=\cases{\Psi_o,&if $0 \le t \le \tau $;\cr
                      0,&otherwise,\cr}\eqno(6)$$
or equivalently,
$$\Psi(t-t')=\cases{\Psi_o,&if $t-\tau \le t' \le t $;\cr
                      0,&otherwise.\cr}\eqno(7)$$
In this case, equation (2) reduces to
$$\eta_{i,k}(t)=\cases{
      >0,&if $[t_{i,k-1},t_{i,k}] \cap [t-\tau,t] \not= \emptyset$;\cr
       0,&otherwise.\cr}\eqno(8)$$
From (8) we see that $\eta_{i,k}(t)$ can be $=0$ depending on the value of
$\tau$ and on the value of $t$ chosen to sample the stellar population.
The most extreme case is that of the SSP ($\tau=0$), for which
$\Psi(t)=\delta(t)$, and $\eta_{i,k}(t)$ in (8) is identical to $N_{i,k}(t)$
in (1).
For a typical set of evolutionary tracks, there is no grid of values $t_j$
of the time variable $t$ for which all the stellar evolutionary stages included
in the tracks can be adequately sampled for arbitrarily chosen values of $\tau$.
This is obviously an undesirable property of population synthesis models.
The models should be capable of representing galaxy properties in a continuous
and well behaved form, independent of the sampling time scale $t$.
Consequently, standard population synthesis models for truncated $\Psi(t)$ as
given by (6)
will reflect the coarseness of the set of evolutionary tracks, and may miss
stellar evolutionary phases depending on our choice of $t$.
This results in unwanted and unrealistic numerical noise in the predicted
properties of the stellar systems studied with the synthesis code
(see example in CB91).

A solution to this problem is to build a set of evolutionary tracks with
a resolution in mass which is high enough to guarantee that all evolutionary
stages, i.e. all values of $k$ in (2), can be populated for any choice of
$t$ or $\tau$ in (6). In other words, there must be tracks for so many stellar
masses in this ideal library that for any model age, we have at hand the
position in the H-R diagram of the star which at that age is in the $k^{th}$
evolutionary stage. The realization of this library is not possible with
present day computers. This limitation can be circumvented by careful
interpolation in a relatively complete set of evolutionary tracks.

\section{The Isochrone Synthesis Algorithm}
The isochrone synthesis algorithm described in this section allows us
to compute continuous isochrones of any age from a carefully selected
set of evolutionary tracks. The isochrones are then used to build
population synthesis models for arbitrary SFRs $\Psi(t)$, without
encountering any of the problems mentioned above. This algorithm
has been used by CB91, BC93, and BC2000. The details of the algorithm follow.

A relationship is built between the main sequence (MS)
mass of a star $M$ and the age $t$ of the star during the $k^{th}$
evolutionary stage (CB91). Ignoring mass loss is justified because
$M$ is used only to label the tracks. For the set of tracks used by
BC93 (described in the next section) 311 different relationships
$log\ M\ vs.\ log\ t$ are built, which can be visualized as 311
different curves in the $(log\ t,\ log\ M)$ plane.
We derive by linear interpolation the MS mass $m_k(t')$ of the star which
will be at the $k^{th}$ evolutionary stage at age $t'$, given by
$$log\ m_k(t')=A_{k,i}\ log(M_i)+(1-A_{k,i})\ log(M_{i+1}),\eqno(9)$$
where
$$A_{k,i}={log\ t_{i+1,k}\ -\ log\ t' \over log\ t_{i+1,k}\ -\ log\ t_{i,k}}.
\eqno(10)$$
$t_{i,k}$ represents the age of the star of mass $M_i$ at the $k^{th}$
evolutionary stage, and
$$t_{i,k} \le t'< t_{i+1,k},$$
and
$$M_{i+1} \le m_k(t') < M_{i}.$$
The procedure is performed for all the curves that intersect the
$log\ t=log\ t'$ line.
We thus obtain a series of values of $m_k$ which must now be
assigned values of $log\ L$ and $log\ T_{eff}$ in order to define the
isochrone corresponding to age $t'$.

To compute the integrated properties of the stellar population, we 
must specify the number of stars of mass $m_k$. This number is determined
from the IMF, which we can write generically as
$$n(m_k)=\phi(m_k^-,m_k^+,1+x).\eqno(11)$$
Of this number of stars,
$$N_{i,k}=A_{k,i}n(m_k)\eqno(12)$$
stars are assigned the observational properties of the star of mass $M_i$
at the $k^{th}$ evolutionary stage, and
$$N_{i+1,k}=(1-A_{k,i})n(m_k)\eqno(13)$$
stars, the observational properties of the star of mass $M_{i+1}$ at the
same $k^{th}$ stage. This procedure is equivalent to interpolating the 
tracks for the stars of mass $M_i$ and $M_{i+1}$ to derive the
values of $log\ L$ and $log\ T_{eff}$ to be assigned to the star of mass
$m_k$, but this intermediate step is unnecessary. In (11) $m_k^-$ and $m_k^+$
are computed from
$$m_k^- = (m_{k-1}m_k)^{1/2},\ \ \ \ \ \ m_k^+ = (m_km_{k+1})^{1/2}.\eqno(14)$$
In this case $m_{k-1},\ m_k$, and $m_{k+1}$ represent the masses obtained by 
interpolation at the given age in the segments representing the $(k-1)^{th},
\ k^{th}$, and $(k+1)^{th}$ stages, respectively.
The IMF has been assumed to be a power law of the form (Salpeter 1955)
$$\phi(m_1,m_2,1+x) = c\int_{m_1}^{m_2}m^{-(1+x)}dm.\eqno(15)$$

For a given $\Psi(t)$ equation (2) gives the number $\eta_{i,k}(t)$ of
stars of mass $M_i$ at the $k^{th}$ evolutionary stage.
If $f_{i,k}(\lambda)$ represents the SED
corresponding to the star of
mass $M_i$ during the $k^{th}$ stage, then the contribution of the 
$\eta_{i,k}(t)$ stars to the integrated evolving SED of the stellar population
is simply given by
$$F_{i,k}(\lambda,t)=\eta_{i,k}(t)f_{i,k}(\lambda).\eqno(16)$$
The resulting SED for the stellar population is given by
$$F(\lambda,t)=\sum_{i,k}F_{i,k}(\lambda,t).\eqno(17)$$

\section{Evolutionary Population Synthesis Models}
A number of groups has developed in recent years different population synthesis
models which provide a sound framework to investigate the problem of spectral
evolution of galaxies.
Some of the most commonly used models are Arimoto \&
Yoshii (1987), Guiderdoni \& Rocca-Volmerange (1987), Buzzoni (1989),
Bressan, Chiosi \& Fagotto (1994), Fritze-v.Alvensleben \& Gerhard (1993),
Worthey (1994), Bruzual \& Charlot (1993, 2000).
The basic astrophysical ingredients used in these models are:
{\it (1)} Stellar evolutionary tracks of one or more metallicities;
{\it (2)} Spectral libraries, either empirical or theoretical model atmospheres;
{\it (3)} Sets of rules, or calibration tables, to transform the theoretical
HR diagram to observational quantities (e.g. $B-V\ vs.\ T_{eff},\ V-K\ vs.
\ T_{eff},\ B.C.\ vs.\ T_{eff}$, etc.). These rules are not necessary when
theoretical model atmosphere libraries are used which are already parameterized 
according to $T_{eff},\ log\ g$, and [Fe/H];
{\it (4)} Additional information, such as analytical fitting functions, 
required to compute various line strength indices (Worthey et al. 1994).
Regardless of the specific computational algorithm used, all evolutionary
synthesis models depend on three adjustable parametric functions:
{\it (1)} the stellar initial mass function, $f(m)$, or IMF;
{\it (2)} the star formation rate, $\Psi(t)$; and
{\it (3)} the chemical enrichment law, $Z(t)$.
For a given choice of $f(m),\ \Psi(t)$, and $Z(t)$, a particular set of 
evolutionary synthesis models provides:
{\it (1)} Galaxy spectral energy distribution $vs.$ time,
$F_\lambda(\lambda,Z(t),t)$;
{\it (2)} Galaxy colors and magnitude $vs.$ time;
{\it (3)} Line strength and other spectral indices $vs.$ time.
Some authors (e.g. Bressan et al. 1994; Fritze-v.Alvensleben \& Gerhard 1994)
consider that $Z(t)$ can be derived self-consistently from their models.
In other instances $Z(t)$ is introduced as an external piece of information.
I discuss below some results from work still in progress in collaboration with
S. Charlot.

\section{Stellar Ingredients}

BC2000 have extended the BC93 evolutionary population synthesis models to
provide the evolution in time of the spectrophotometric properties of 
SSPs for a wide range of stellar metallicity.
In an SSP all the stars form at $t = 0$ and evolve passively afterward.
The BC2000 models are based on the stellar evolutionary tracks
computed by Alongi et al. (1993), Bressan et al. (1993), Fagotto et al.
(1994a, b, c), and Girardi et al. (1996), which use the radiative opacities
of Iglesias et al. (1992).
This library includes tracks for stars with initial chemical composition
$Z = 0.0001, 0.0004, 0.004, 0.008, 0.02, 0.05$, and 0.10 (Table 1),
with $Y= 2.5Z + 0.23$, and initial mass $0.6 \leq m/M_\odot \leq 120$ for all
metallicities, except $Z=0.0001$ ($0.6 \leq m/M_\odot \leq 100$)
and $Z=0.1$ ($0.6 \leq m/M_\odot \leq 9$).
This set of tracks will be referred to as the Padova or $P$ tracks hereafter.
A similar set of tracks for slightly different values of $Z$ has been
published by Girardi et al. (2000). A comparison of the
predictions of models built with both sets of Padova tracks will be
shown elsewhere (but see Fig. 7 and Bruzual 2000).

\begin{figure}
\plotfiddle{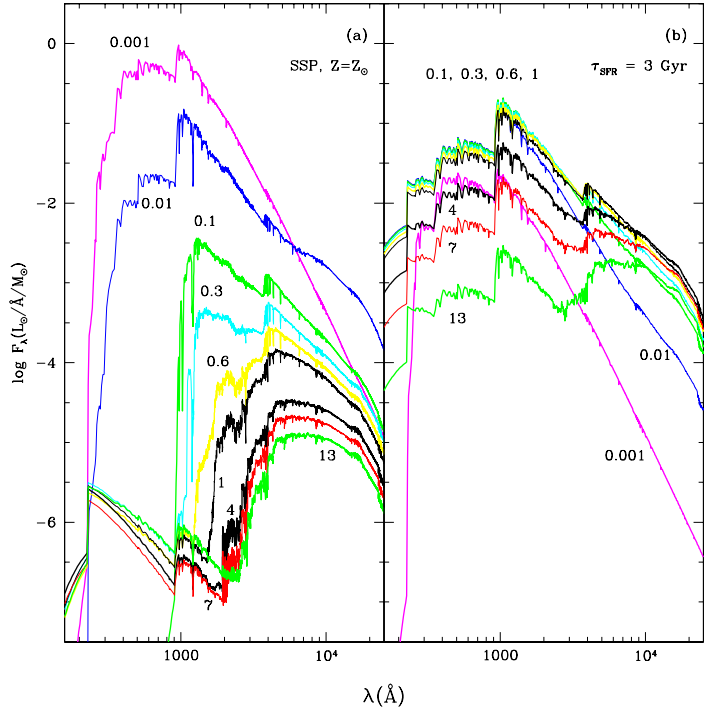}{10cm}{0}{50}{50}{-150}{-80}
\caption{
Evolving spectral energy distributions. (a) Evolution in time of the SED of a
SSP computed for the Salpeter IMF ($m_L = 0.1,\ m_U = 125~M_\odot)$.
The age in Gyr is indicated next to each spectrum.
(b) Same as (a) but for a composite population in which stars
form according to $\Psi(t)=\exp(-t/\tau)$ for $\tau=3$ Gyr.
The total mass of each model galaxy is 1 $M_\odot$. $F_\lambda$ in
frame (b) has been multiplied by 100 to use a common vertical scale.
}
\end{figure}

The published tracks go through all phases of stellar evolution
from the zero-age
main sequence to the beginning of the thermally pulsing regime of the
asymptotic giant branch (AGB, for low- and intermediate-mass stars)
and core-carbon ignition (for massive stars), and include mild overshooting
in the convective core of stars more massive than $1\ M_{\odot}$.
The Post-AGB evolutionary phases for low- and intermediate-mass stars
were added to the tracks by BC2000 from different sources
(see BC2000 for details).

BC2000 use as well a parallel set of tracks for solar metallicity
computed by the Geneva group (Geneva or $G$ tracks hereafter), which
provides a framework for comparing models computed with two
different sets of tracks.

The BC2000 models use the library of synthetic stellar spectra compiled
by Lejeune et al. (1997, 1998, LCB97 and LCB98 hereafter) for all the
metallicities in Table 1.
This library consists of Kurucz (1995) spectra for the hotter stars (O-K),
Bessell et al. (1989, 1991) and Fluks et al. (1994) spectra for M giants,
and Allard \& Hauschildt (1995) spectra for M dwarfs.
For $Z = Z_\odot$, BC2000 also use the Pickles (1998) stellar atlas,
assembled from empirical stellar data.

\begin{table}
  \begin{center}
   \begin{minipage}{6cm}
    \begin{tabular}{cccc}
   Z    &    X    &    Y   &  [Fe/H] \\ \hline
0.0001  & 0.7696  & 0.2303 & -2.2490 \\
0.0004  & 0.7686  & 0.2310 & -1.6464 \\
0.0040  & 0.7560  & 0.2400 & -0.6392 \\
0.0080  & 0.7420  & 0.2500 & -0.3300 \\
0.0200  & 0.7000  & 0.2800 &  0.0932 \\
0.0500  & 0.5980  & 0.3520 &  0.5595 \\
0.1000  & 0.4250  & 0.4750 &  1.0089 \\ \hline
    \end{tabular}
   \end{minipage}
  \end{center}
\caption{Model chemical composition}\label {}
\end{table}

\section{Spectral evolution at fixed metallicity}

\begin{figure}
\plotfiddle{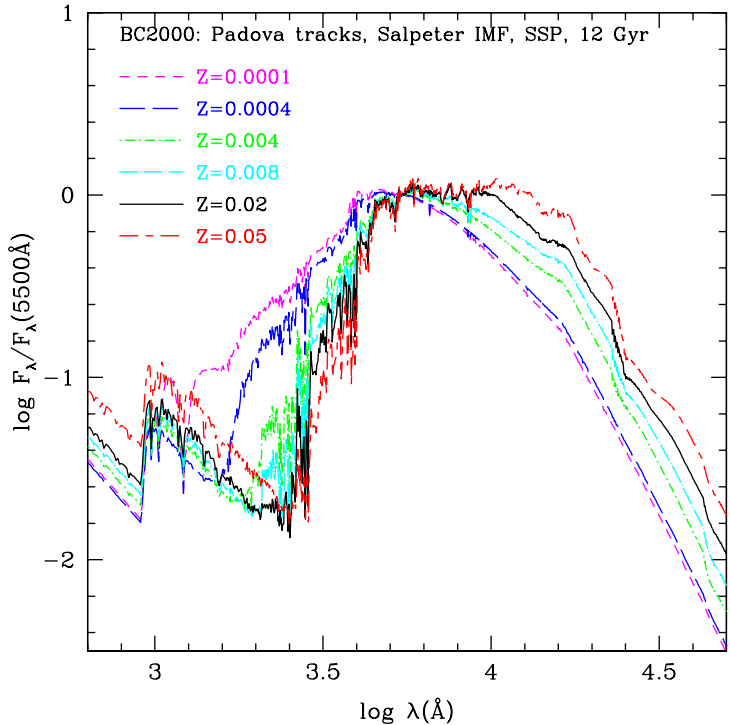}{10cm}{0}{50}{50}{-150}{-80}
\caption{
Chemically homogeneous BC2000 SSP model galaxy SEDs at age = 12 Gyr.
Each line pattern represents a different metallicity, as indicated inside the
frame. All the models shown were computed for the Salpeter
IMF ($m_L = 0.1,\ m_U = 125~M_\odot)$.
The total mass of each model galaxy is 1 $M_\odot$.
The SEDs have been normalized at $\lambda = 5500$\AA.
}
\end{figure}

\begin{figure}
\plotfiddle{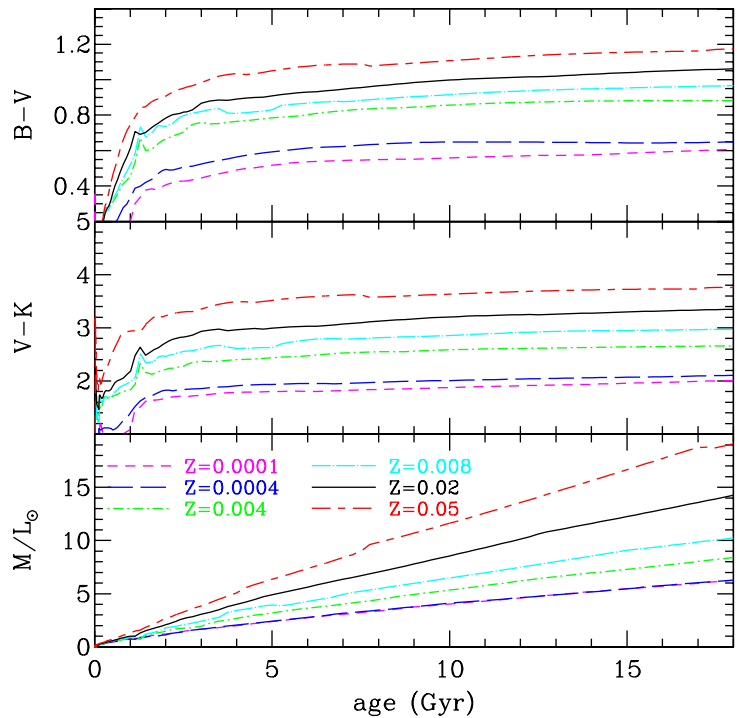}{10cm}{0}{50}{50}{-150}{-80}
\caption{Evolution in time of the $B-V$, and $V-K$ colors, and
the $M/L_V$ ratio for the BC2000 SSP models shown in Fig. 2.
Each line represents a different metallicity, as indicated in the bottom panel.}
\end{figure}

\begin{figure}
\plotfiddle{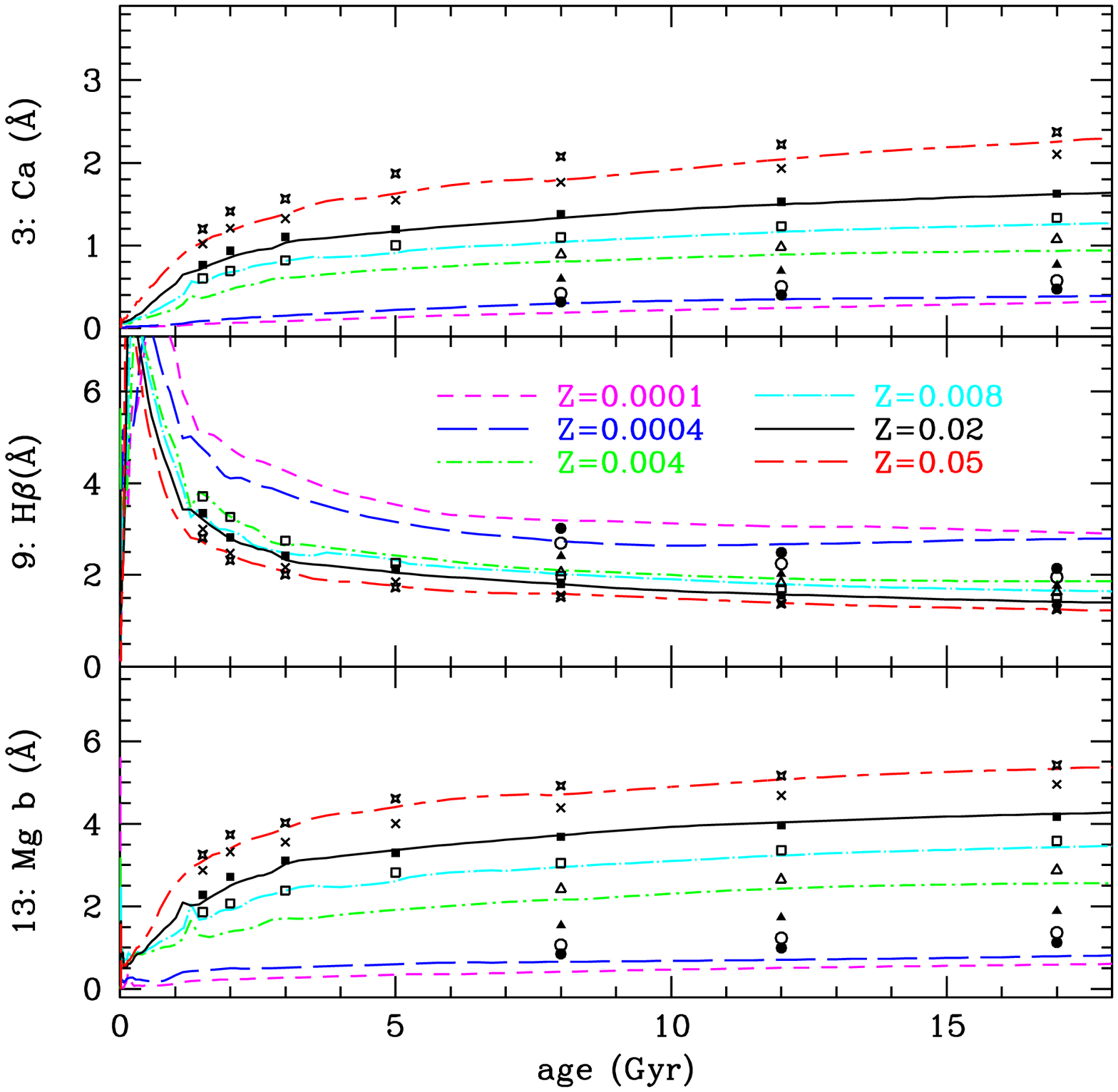}{10cm}{0}{50}{50}{-150}{-80}
\caption{Evolution in time of the Mg$_b$, H$_\beta$, and Ca spectral indices
as defined by Worthey (1994) for the BC2000 models shown in Fig. 2.
The different symbols represent the values of the indices computed by
Worthey for the same range in $Z$. Each line represents a different
metallicity, as indicated in the middle panel.}
\end{figure}

Fig. 1a shows the evolution in time of the SED for the SSP model.
In an SSP all the stars form at $t = 0$ and evolve passively afterward.
In all the examples shown in this paper I assume that stars form according
to the Salpeter (1955) IMF in the range from $m_L = 0.1$ to $m_U = 125~M_\odot$.
The total mass of the model galaxy is 1 $M_\odot$.
The evolution is fast and is dominated by massive stars during the first
Gyr in the life of the SSP (6 top SEDs). The flux seen around 2000 {\AA}
at 4 and 7 Gyr is produced by the turn-off stars. The UV-rising branch
(Burstein {\it et al.} 1988, Greggio and Renzini 1990) seen after 10 Gyr is 
produced by the PAGB stars. These stars are
also responsible of the decrease in the amplitude of the 912 {\AA} discontinuity
observed after 4 Gyr.
The SSP model is the basic ingredient which, together with the convolution
integral (2), is used to compute models with arbitrary SFRs and equal IMF.
For illustration I show in Fig. 1b the evolution
of a model with $\Psi(t)=\exp(-t/\tau)$ for $\tau=3$ Gyr.
The UV to optical spectrum remains roughly constant during the main
episode of star
formation because of the continuous input of young massive stars, but the
near-infrared light rises as evolved stars accumulate. When star formation
drops, the spectral characteristics at various wavelengths are determined by
stars in advanced stages of stellar evolution.

\section{Dependence of galaxy properties on stellar metallicity}

Fig. 2 shows the predicted SEDs at $t = 12$ Gyr for chemically homogeneous
SSPs of the indicated metallicity.
The SEDs shown in Fig. 2 have been normalized at $\lambda = 5500$\AA\ to make
the comparison more clear. 
Fig. 3. shows the evolution in time of the $B-V$ and $V-K$ colors, and the 
$M/L_V$ ratio predicted by BC2000 for the same SSPs shown in Fig. 2.

From Figs. 2 and 3 it is apparent that there is a uniform tendency for
galaxies to become redder in $B-V$ as the metallicity increases from 
$Z = 0.0001~({1 \over 200} Z_\odot)$ to $Z = 0.05~(2.5 \times Z_\odot)$.
The $V-K$ color and the $M/L_V$ ratio show the expected
tendency with metallicity, i.e. $V-K$ becomes redder and $M/L_V$ becomes
higher with increasing $Z$.

Fig. 4 shows the evolution in time of the Mg$_b$, H$_\beta$, and Ca spectral
indices as defined by Worthey (1994) for the same BC2000 SSP models shown
in Figs. 2 and 3. Again,
the models show the
expected tendency with $Z$ and match the values computed by Worthey (1994).
It should be remarked that the time behavior of the line strength indices at
constant $Z$ is due to the change in the number of stars at different
positions in the HR diagram produced by stellar evolution and is not
related to chemical evolution. The indices change also in chemically
homogeneous populations.
The H$_\beta$ index is less sensitive to the stellar metallicity than the
Mg$_b$ and Ca index. Instead, the H$_\beta$ index is high when there is a large
fraction of MS A-type stars ($t < 1$) Gyr.

\begin{figure}
\plotfiddle{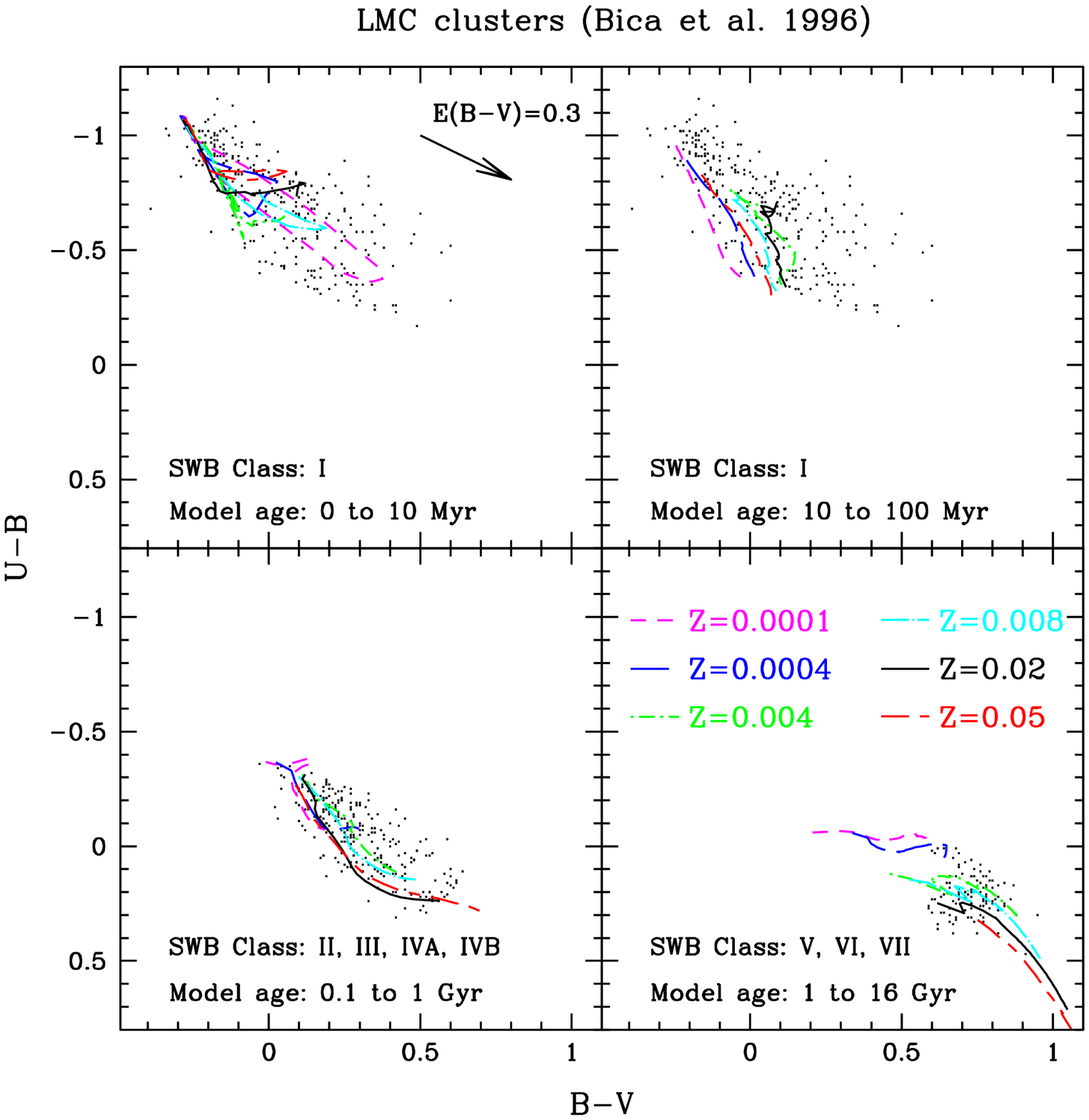}{10cm}{0}{50}{50}{-150}{-80}
\caption{
Points: LMC cluster $(U-B)$ vs. $(B-V)$ colors from Bica et al. (1996b),
discriminated by SWB class. The lines represent SSP models at various
age ranges.}
\end{figure}

In Fig. 5 I compare the behavior of the SSP models in the $(U-B)$ vs. $(B-V)$
color plane with the LMC cluster data from Bica et al. (1996b), discriminated
by SWB class and model age. In each panel the models (lines) are shown in
the range of age for which the predicted colors overlap the observed colors
for the class.
It is apparent from the figure that the models reproduce quite well the
observed colors.

\begin{figure}
\plotfiddle{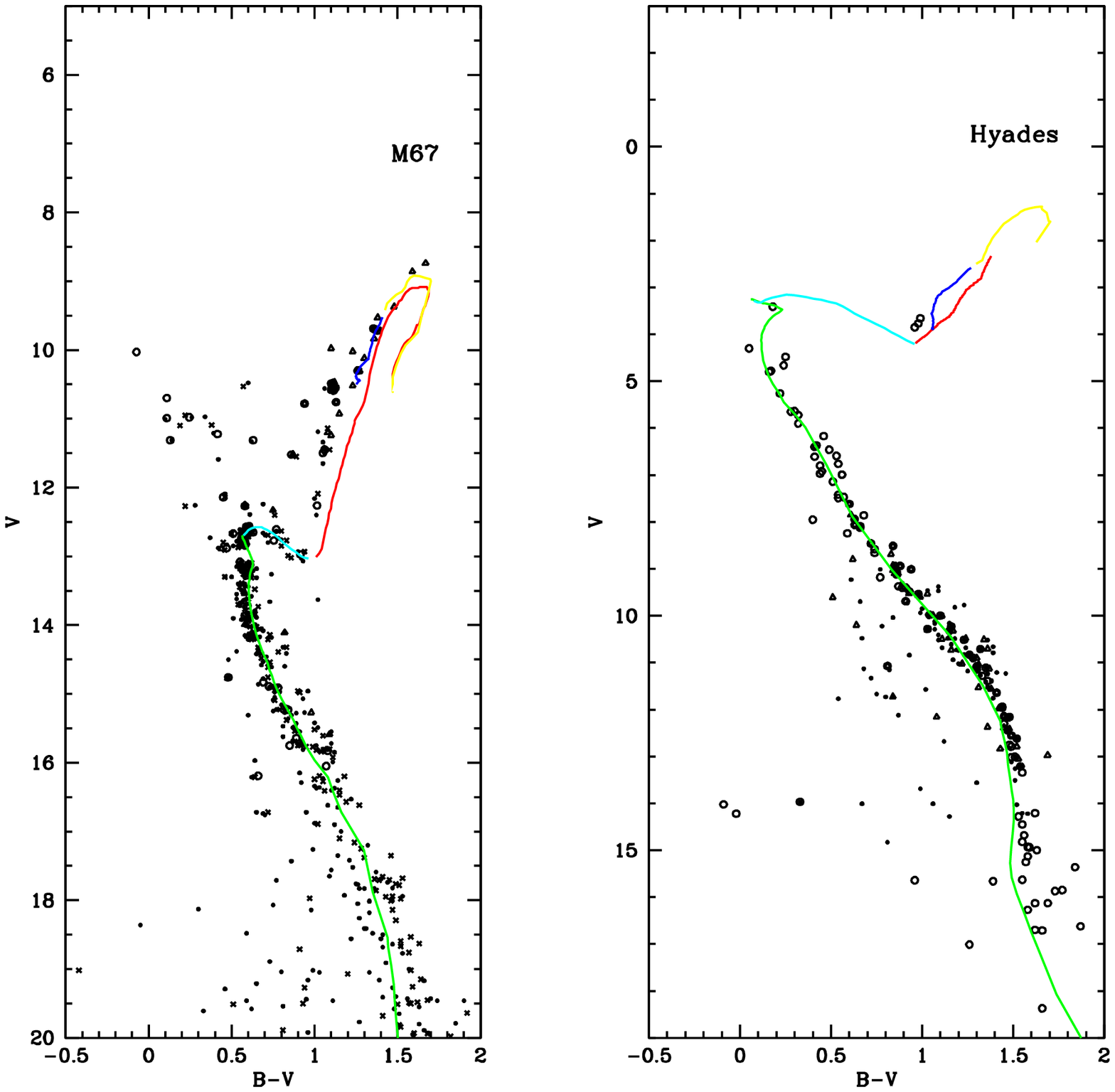}{10cm}{0}{50}{50}{-150}{-80}
\caption{
CMD of M67 and the Hyades compared with isochrones derived from the Padova
tracks for solar metallicity. See text for details.
}
\end{figure}


\begin{figure}
\plotfiddle{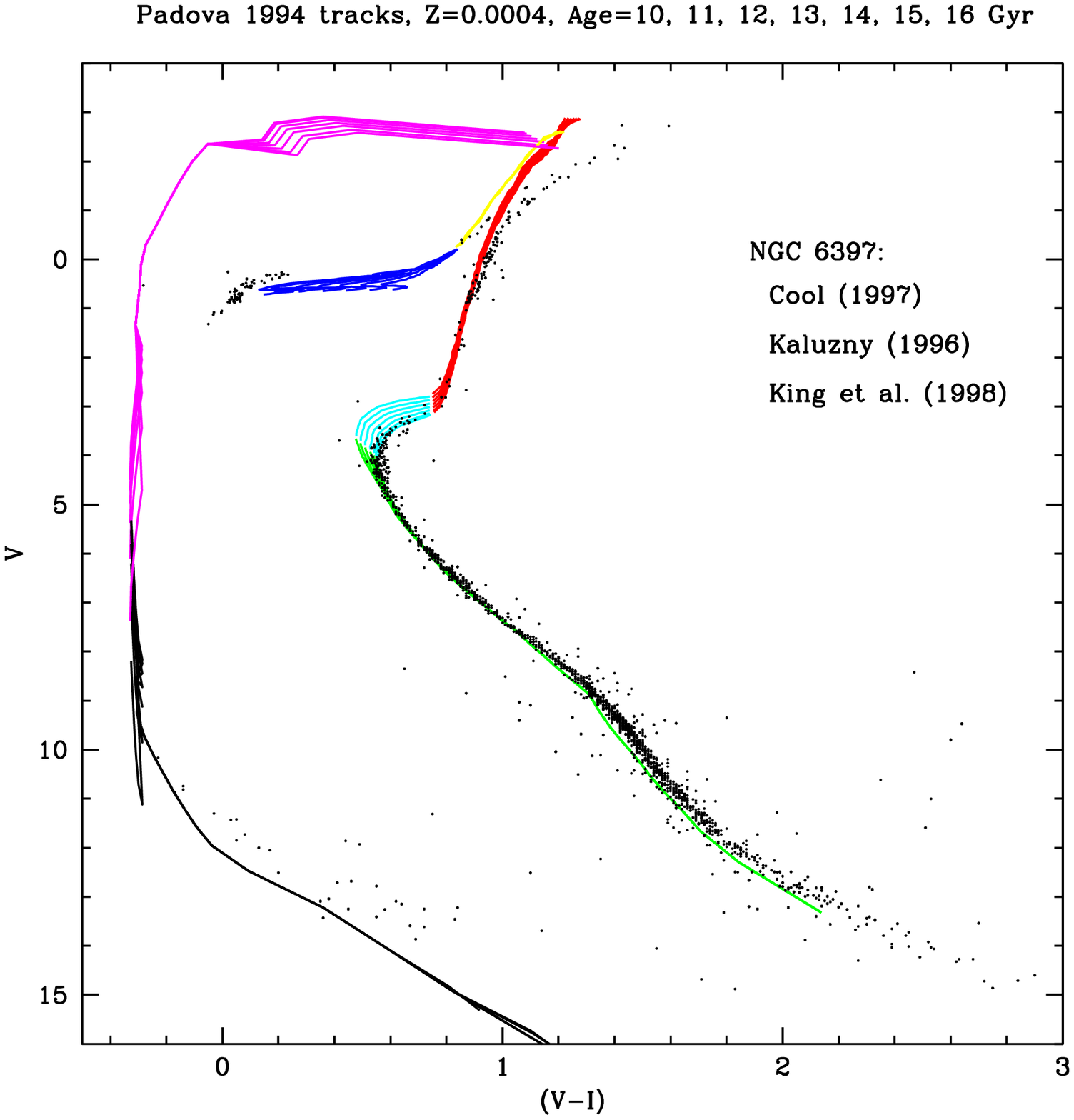}{10cm}{0}{50}{50}{-150}{-80}
\plotfiddle{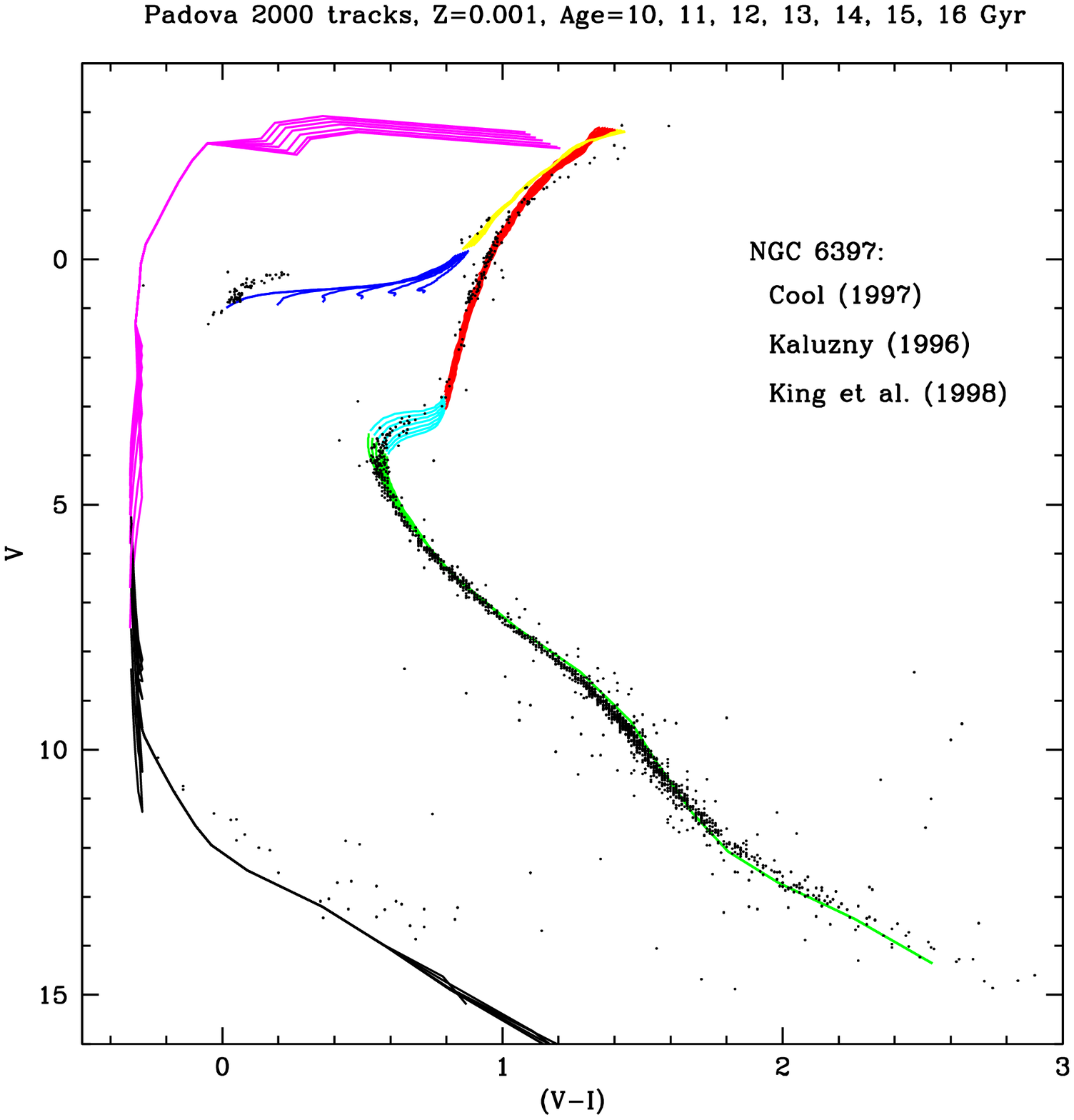}{10cm}{0}{50}{50}{-150}{-80}
\caption{
CMD of NGC 6397 compared with isochrones derived from
the Padova-1994 tracks for $Z=0.0004$ (top), and
the Padova-2000 tracks for $Z=0.001$ (bottom).
The observations were assembled by D'Antona (1999)
from the sources indicated in the figure.
}
\end{figure}

\section{Calibration of the models in the C-M diagram}

Population synthesis models, frequently used to study composite
stellar populations in distant galaxies, are rarely confronted with
observations of local simple stellar populations, such as globular star
clusters, whose age and metal content have been constrained considerably
in later years. If the models cannot reproduce the CMDs and SEDs of
these objects for the correct choice of parameters,
their predictive power becomes weaker and their usage to study complex
galaxies may not be justified.

It is important to compare the properties of the population synthesis
models to observations of stellar systems whose age and metallicity is well
determined.
This is a means to test to what extent the adopted relationships between
stellar color and magnitude and effective temperature and luminosity (or
surface gravity) introduce systematic shifts between the predicted
and observed isochrones in the C-M diagram.

For each value of the metallicity $Z$ listed in Table 1,
and a particular choice of the IMF $\phi(m)$,
a BC2000 SSP model consists of a set of
221 evolving integrated SEDs spanning from 0 to 20 Gyr.
The isochrone synthesis algorithm used to build these models renders it
straightforward to compute the loci described by the stellar population in
the CMD at any time step and in any photometric band.
We can extract the model SED that best reproduces a given observed
SED and assign an age to the program object, and then examine how
well the isochrone computed at this age fits
the most significant features in the CMD of this object, if available.

Fig. 6 compares the $(B,B-V)$ observations of the clusters M67 and the
Hyades with the isochrones obtained from the BC2000 models.
These two clusters have nearly solar metallicity: M67
([Fe/H]$\approx0.01$), the Hyades ([Fe/H]$\approx0.15$).
For M67 we adopted a distance modulus of 9.5 mag and a color
excess $E(B-V)=0.06$ mag (Janes 1985). For the Hyades a distance modulus of 3.4
mag (Peterson and Solensky 1988) and $E(B-V)\approx0$.
Estimates of the ages of these clusters vary from 4 to 4.3 Gyr for M67
and from 0.5 to 0.8 Gyr for the Hyades.
The isochrones are shown at 4 Gyr (M67) and 0.6 Gyr (Hyades).
The data points for M67 are from Eggen and Sandage (1964, {\it open circles}),
Racine (1971, {\it filled circles}), Janes and Smith (1984, {\it triangles}),
and Gilliland et al. (1991; {\it crosses}).
For the Hyades the observations are from Upgren (1974, {\it triangles}),
Upgren and Weis (1977, {\it filled circles}), and Micela et al. (1988, {\it open
circles}).

Fig. 7 shows a comparison of the excellent HST CMD diagram of NGC 6397
assembled from various sources by D'Antona (1999) with isochrones computed
from the Padova-1994 tracks for $Z = 0.0004$ and the Padova-2000 tracks
for $Z=0.001$ at ages 10 to 16 Gyr.
The $original$ version of the model atmospheres in the LCB atlas was
used to derive the colors in Fig. 7. The $corrected$ version of these
models produces considerably worse agreement with the observations, mainly
in the MS from the turn-off down.
The redder cluster RGB most likely reflects a slightly higher metallicity
than $Z=0.0004$, close to $Z=0.001$.
Despite the discrepancies seen in Figs. 6 and 7 (mainly in the RGB of M67 and
NGC 6397) the agreement between the predicted isochrones and the loci in the
CMD of these clusters may be regarded as satisfactory, and is excellent in
some parts of the diagram.

\begin{figure}
\plotfiddle{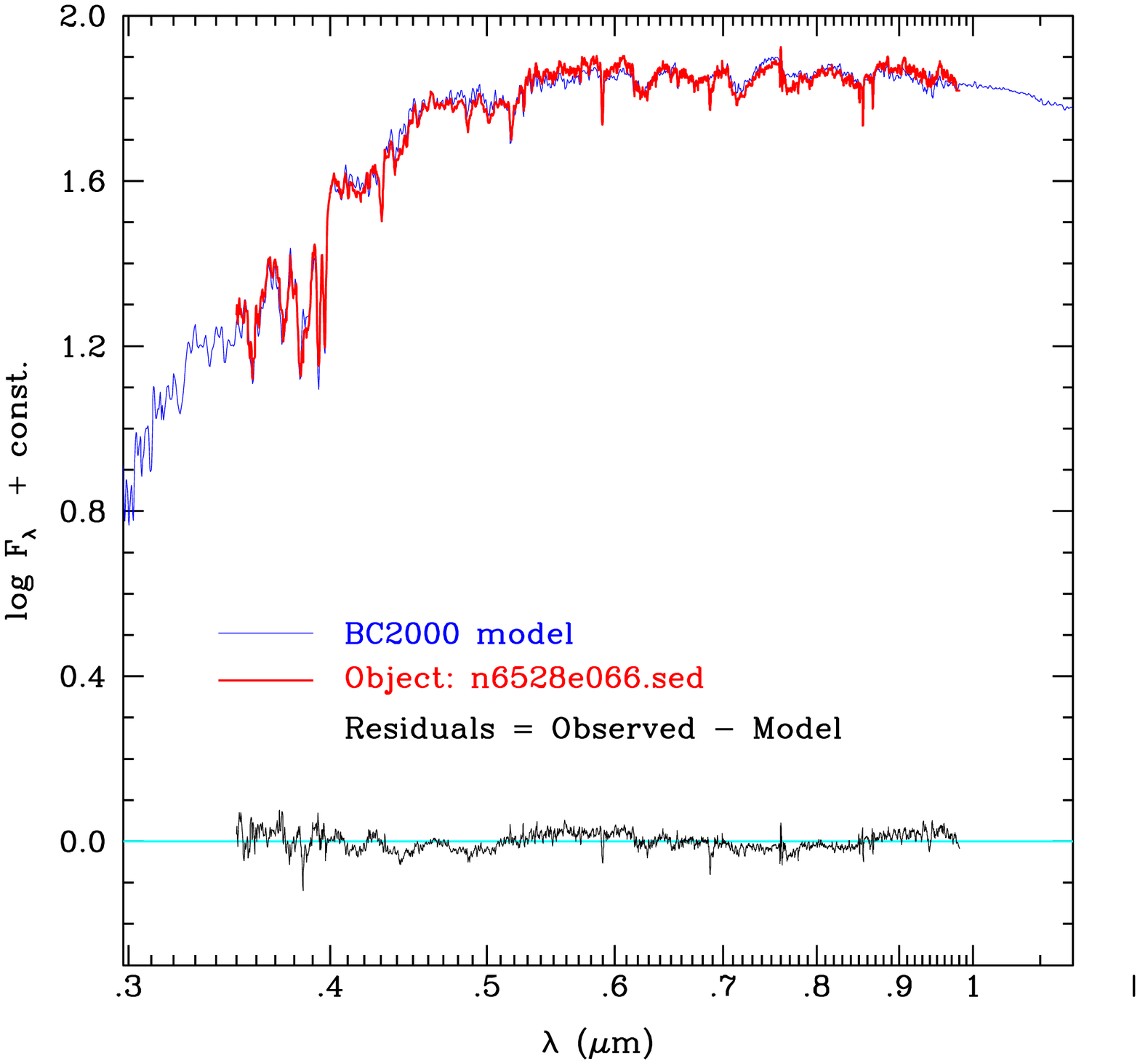}{10cm}{0}{50}{50}{-150}{-80}
\caption{
Best fit to the integrated spectrum of NGC 6528 (heavy line) in the
range $\lambda\lambda$ 3500 - 9800 ${\rm \AA}$ for model 2 in Table 2
(thin line extending over the full wavelength range).
The best fit occurs at 10.25 Gyr.
The residuals of the fit, $log~ F_\lambda(observed) - log~ F_\lambda(model)$,
are shown as a function of wavelength.}

\plotfiddle{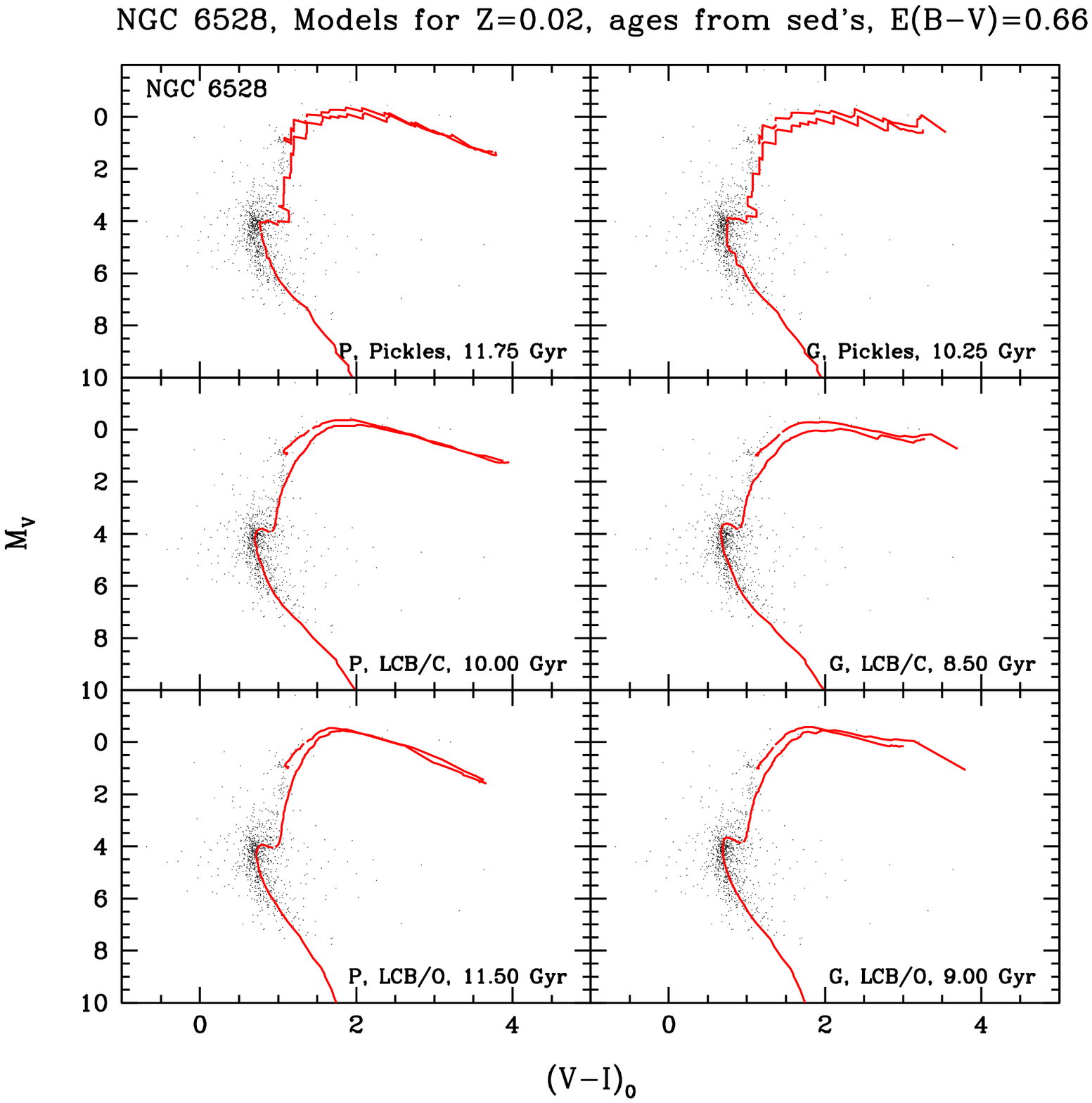}{10cm}{0}{50}{50}{-150}{-80}
\caption{
Intrinsic $M_V$ vs. $(V-I)_0$ CMD of NGC 6528 shown together with
theoretical isochrones for the $Z = Z_\odot$ models listed in Table 2.}
\end{figure}

\section{Observed Color-Magnitude Diagrams and Integrated Spectra}

The high quality and depth of the HST VI-photometry NGC 6528
($Z \simlt Z_\odot$), as well as the high signal-to-noise ratio of the
integrated SED of NGC 6528 currently available, provide an excellent
framework for testing the range of validity of population synthesis models
(see Bruzual et al. 1997 for details).

The integrated spectrum of NGC 6528 in the wavelength range
$\lambda =$ 3500 - 9800 ${\rm \AA}$ was obtained by combining
the visible, near-infrared, and near-ultraviolet spectra of 
Bica \& Alloin (1986, 1987) and Bica et al. (1994), respectively.
We applied a reddening correction of E(B-V) = 0.66, adopted in Bica et al.
(1994 and references therein).
The SED of NGC 6528 is typical of old stellar populations of high metal
content (Santos et al. 1995). 
It may happen that NGC 6528, similarly to NGC 6553, has [Fe/H] $<$ 0.0
(Barbuy et al. 1997a), whereas the [$\alpha$-elements/Fe] are enhanced,
resulting in [Z/Z$_{\odot}$] $\approx$ 0.0, or possibly slightly below solar.
Since we have evolutionary tracks for $Z = 0.02$ and $Z = 0.008$,
we adopt $Z = 0.02$.

\begin{table}
  \begin{center}
   \begin{minipage}{8cm}
    \begin{tabular}{cccrr}
       &Spectral & Stellar & Best-fitting &                   \\ 
Model  &Library  & Tracks  & age (Gyr)    & $\Sigma^2_{min}$  \\ \hline
  1    & Pickles &   $P$   & 11.75        &   2.22            \\
  2    &  "      &   $G$   & 10.25        &   1.34            \\
       &         &         &              &                   \\
  3    & LCB97-C &   $P$   & 10.00        &   1.73            \\
  4    &  "      &   $G$   &  8.50        &   1.38            \\
       &         &         &              &                   \\
  5    & LCB97-O &   $P$   & 11.50        &   3.35            \\
  6    &  "      &   $G$   &  9.00        &   1.95            \\ \hline
    \end{tabular}
   \end{minipage}
  \end{center}
\caption{$Z_\odot$ model fits to SED of NGC 6528}
\end{table}

The data available for this cluster provide a unique
opportunity to examine the 6 different options for the $Z = Z_\odot$ models
considered by BC2000.
Thus, we can study objectively which of the basic building blocks used by
BC2000 for $Z = Z_\odot$: $P$ or $G$ tracks; Pickles,
LCB97-O (original version),
or LCB97-C (corrected version) stellar libraries, is most successful at
reproducing the data.
In Table 2 I show the age at which $\Sigma^2$, defined as the sum of squared
residuals $[ log~ F_\lambda(observed) - log~ F_\lambda(model)]^2$, is minimum
for various $Z = Z_\odot$ models.
The values of $\Sigma^2_{min}$ given in Table 2 indicate the goodness-of-fit.
According to this criterion, models 2 and 4 provide the best fit to
the integrated SED of NGC 6528 in the wavelength range $\lambda$ 3500-9800
${\rm \AA}$. This fit is shown in Fig. 8.
Except for the differences in the best-fitting age, model
3 provides a comparable,
although somewhat poorer, fit to the SED of this cluster.
The residuals for models 1, 5 and 6 are considerably larger.
Spectral evolution is slow at these ages, and the minimum in the function
$\Sigma^2$ vs. age is quite broad.
Reducing or increasing the model age by 1 or 2 Gyr produces fits of
comparable quality to the one at which $\Sigma^2$ is minimum.
For instance, for model 2, $\Sigma^2$(12 Gyr) = 1.88, and
$\Sigma^2$(8 Gyr) = 2, which are still better fits according to the $\Sigma^2$
criterion than the best fits provided by some models in Table 2.

Fig. 9 shows the intrinsic HST VI CMD of NGC 6528 together with
the isochrones corresponding to the models in Table 2.
It is apparent from this figure that all the isochrones shown 
provide a good representation of the cluster population in this CMD,
especially the position of the turn-off and the base of the asymptotic
giant branch (AGB).
We note that NGC 6528 shows a double turnoff, the upper one
being due to contamination from the field star main sequence. The appropriate
TO location would be around that indicated by the 10, 11 and 12 Gyr isochrones.
Despite the fact that models 2 and 4 provide better fits to the SED of this
cluster than models 1 and 3, the isochrones from model 3 reproduce more
closely the CMD diagram than models 1 and 2. The noisy nature of the isochrones
computed with models 1 and 2 is due to the lack of some stellar types in the
Pickles stellar library.

{From} these results we conclude that:

(1) The $\lambda = 3500-9800$ ${\rm \AA}$ SED for E(B-V)=0.66 and the VI CMD of
NGC 6528 are well reproduced by $Z = Z_\odot$ models at an age from 9
to 12 Gyr.

(2) The age derived from the fit to the observed SED of NGC 6528 is
extremely sensitive to the assumed E(B-V). Using E(B-V) = 0.59 instead of
0.66, increases the best-fitting ages by 2 to 3 Gyr, since the observed
SED is then intrinsically redder. On the other hand, if E(B-V) = 0.69,
the observed SED is intrinsically bluer and the derived ages are 2 to 3 Gyr
younger than for E(B-V) = 0.66. However, inspecting the isochrones in the
CMD, the ages derived for E(B-V) = 0.66 listed in Table 2, seem appropriate.

(3) The SED and CMD of this cluster are consistent with those expected for
a $Z = Z_\odot$ population at an age of $\approx$ 9-12 Gyr, if overshooting
occurs in the convective core of stars down to $1\ M_{\odot}$ ($P$ tracks,
models 1, 3, and 5 in Table 2). If overshooting stops at $1.5\ M_{\odot}$,
as in the $G$ tracks, this age is reduced to $\approx$ 8-10 Gyr
(models 2, 4, and 6 in Table 2).

(4) For the same spectral library, the ages derived from the $P$ tracks
are older than the ones derived from the $G$ tracks. This is due to the
fact that the $P$ tracks include overshooting in the convective core of
stars more massive than $1\ M_{\odot}$ whereas the $G$ tracks stop
overshooting at $1.5\ M_{\odot}$. Thus, stars in this mass range require
more time in the $P$ tracks to leave the main sequence than in the $G$ tracks.

(5) For the same set of evolutionary tracks, the corrected LCB97 library seems
to provide a better fit to the CMD than the Pickles atlas.
Interpolation in the
finer LCB97 grid of models produces smoother isochrones than in the coarser
Pickles atlas. We attribute this to the fact that M stars are very sparse in
the Pickles atlas. Furthermore, the temperature scale becomes problematic
for these stars in the Pickles atlas.
In the LCB97 library, the temperature scale
for giants relies on measurements of angular diameters and fluxes, which
enter directly in the definition of effective temperature. For dwarfs,
the temperature scale is more difficult to define, as discussed in LCB98.

(6) Noticeable differences exist in the isochrones computed for both sets of
LCB97 libraries.
The differences are more pronounced for the M giants of
$(V-I)_0 > 1.6$, and $(J-K)_0 > 1$, corresponding to a temperature of
${\rm T_e \leq 4000K}$, and for the cool dwarfs of $(V-I)_0$ $>$ 1,
corresponding to a temperature of ${\rm T_e \leq 4700K}$.
There are very few of these stars in the CMD of NGC 6528 to favor a particular
choice of library. However, the corrected library produces better fits to
the observed SED. We attribute this fact to the relative importance of the
luminosity of M giants.

(7) In general the LCB97 corrections redden the stellar SEDs in the optical
range, producing redder SSP models at an earlier age. As a consequence,
the ages derived in Table 1 for the LCB97-C library are younger than the
ones derived from the LCB97-O library for the same set of tracks.

(8) We have adopted $Z = Z_\odot$ for this cluster. However, for a slightly
lower value of $Z$, the derived age would be older.

\section{Comparison of model and observed spectra}

\begin{figure}
\plotfiddle{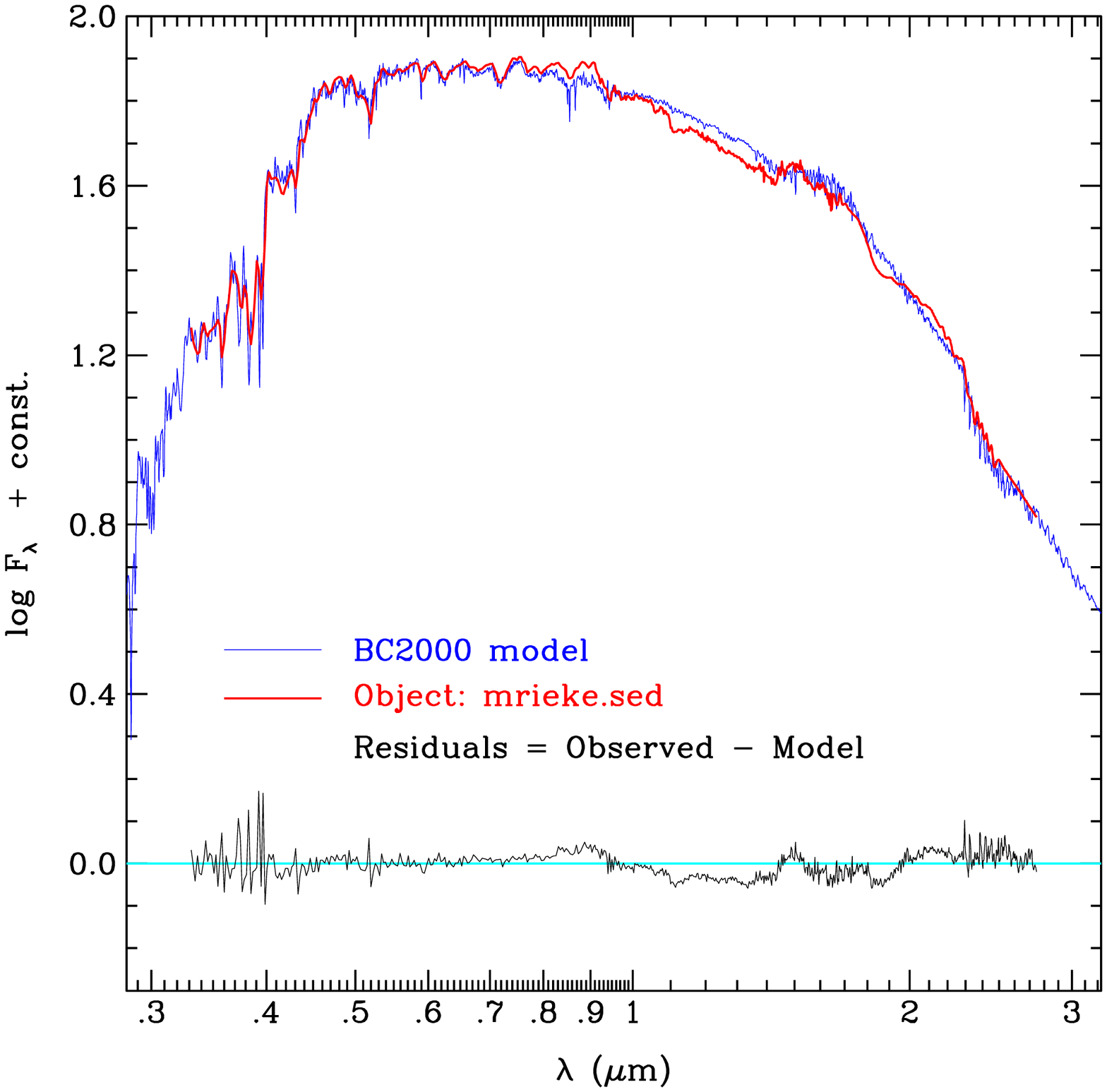}{9cm}{0}{50}{50}{-150}{-80}
\caption{
Best fit to an average Elliptical galaxy SED (heavy line) in the
range $\lambda\lambda$ 3300 - 27500 ${\rm \AA}$. The model is the
thin line extending over the full wavelength range.
The best fit occurs at 10 Gyr for this model SED.
The residuals of the fit, $log~ F_\lambda(observed) - log~ F_\lambda(model)$,
are shown as a function of wavelength.}

\plotfiddle{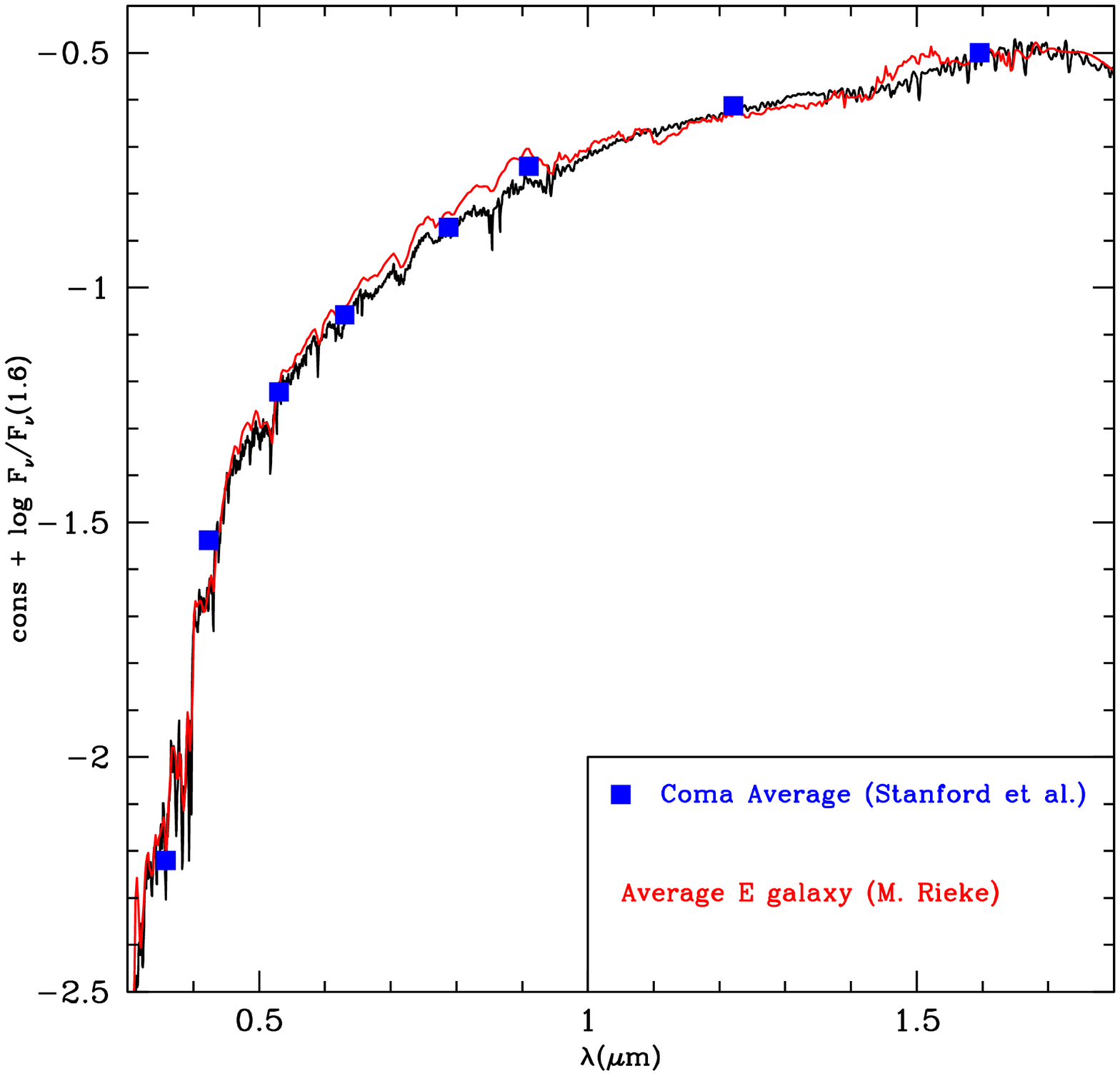}{9cm}{0}{50}{50}{-150}{-80}
\caption{
Best fit to an average Elliptical galaxy SED (same model and E galaxy SED as
in Fig. 10 but in different units). The broad band fluxes representing
the average of many E galaxies in the Coma cluster (solid squares)
from A. Stanford (private communication) are shown.
The observed SED is the one with the lowest spectral resolution.
}
\end{figure}

\begin{figure}
\plotfiddle{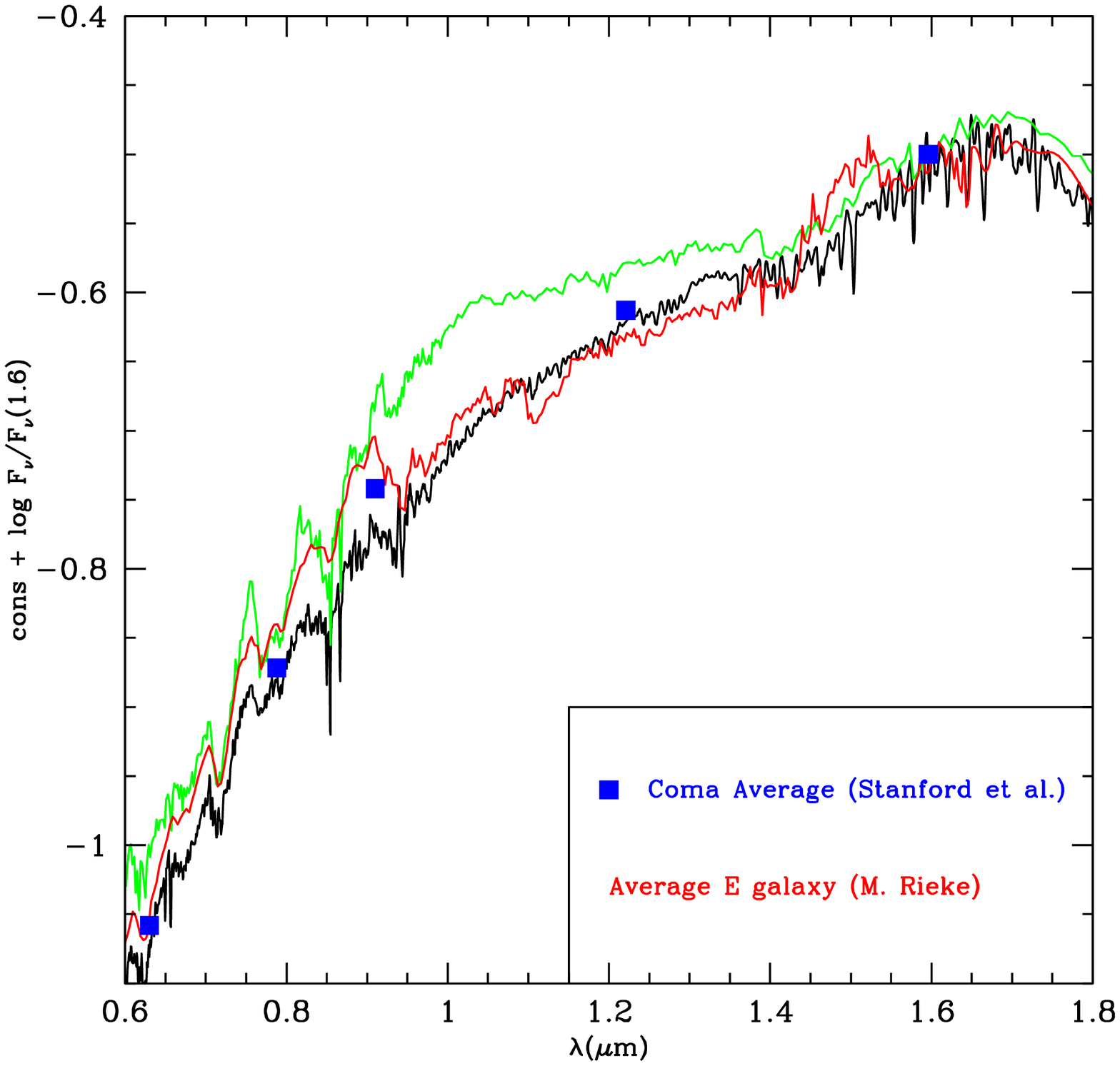}{9cm}{0}{50}{50}{-150}{-80}
\caption{
Fits to an average Elliptical galaxy SED.
This figure shows a closer look at the same data of Fig. 11 in an enlarged scale.
The discrepant line in corresponds to the same model shown in Figs 10 and
11 but using the LCB97 synthetic stellar atlas instead of the empirical
stellar SEDs.
}
\end{figure}

\subsection{Solar metallicity}

Fig. 10 shows a model fit to the average spectrum of an E galaxy (kindly
provided by M. Rieke). The model SED is the line extending over the complete
wavelength range shown in the figure. The observed SED covers the range
from 3300 \AA\ to 2.75 $\mu$m. The residuals (observed - model) are shown
at the bottom of the figure in the same vertical scale. The model 
corresponds to a 10 Gyr $Z = Z_\odot\ $ SSP computed for the Salpeter (1955) IMF
($m_L = 0.1,\ m_U = 125~M_\odot)$ using the $P$ tracks and the
Pickles (1998) stellar atlas. The fit is excellent over most of the
spectral range. A minor discrepancy remains in the region from 1.1 to 1.7
$\mu$m. The source of this discrepancy is not understood at the moment.
In Fig. 11 I show the same model and E galaxy SED as in Fig. 10 but in different
units. In addition, in Fig. 11 I include the broad band fluxes representing
the average of many E galaxies in the Coma cluster (solid squares)
from A. Stanford (private communication).
The observed SED is the one with the lowest spectral resolution.
Fig. 12 shows a closer look at the same data in an enlarged scale.
Again the agreement is excellent for the 3 data sets.
The discrepant line in Fig. 12 corresponds to the same model shown in Figs 10 and
11 but I used the LCB97 synthetic stellar atlas instead of the empirical
stellar SEDs. Fig. 12 shows clearly that the models based on empirical stellar
SEDs are to be preferred over the ones based on theoretical model
atmospheres. Unfortunately, complete libraries of empirical stellar SEDs
are available only for solar metallicity.

\begin{figure}
\plotfiddle{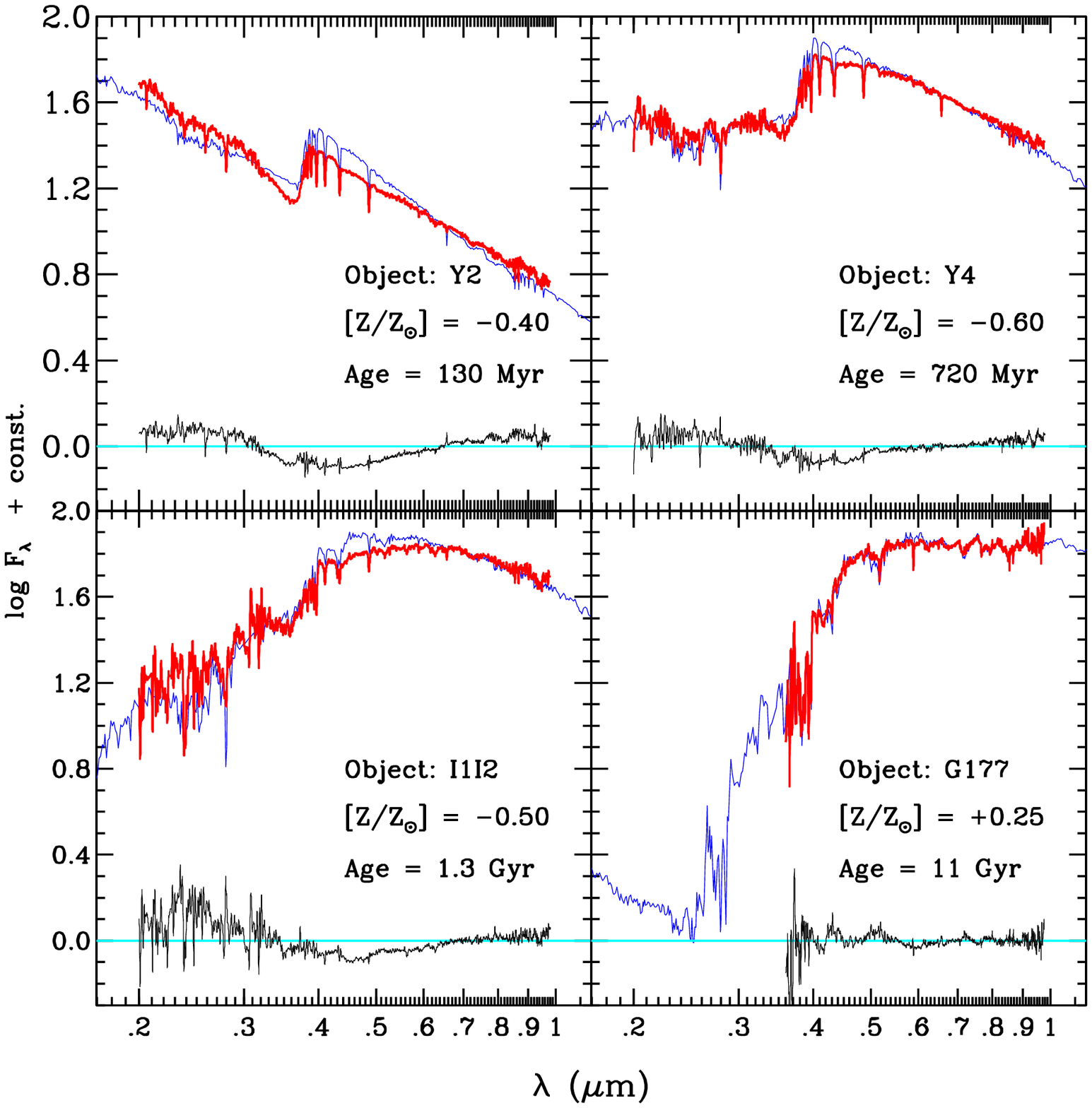}{9cm}{0}{50}{50}{-150}{-80}
\caption{
Best fit to average optical SED (heavy line) of star clusters of various
metallicities compiled by Bica et al. (1996a).
The SSP model is the thin line extending over the full wavelength range.
The residuals of the fit, $log~ F_\lambda(observed) - log~ F_\lambda(model)$,
are shown as a function of wavelength.
The name and the metal content of the observed spectra indicated in
each panel is as given by Bica et al. The quoted age is derived from
the best fit of our model spectra for the indicated metallicity
to the corresponding observations.
}

\plotfiddle{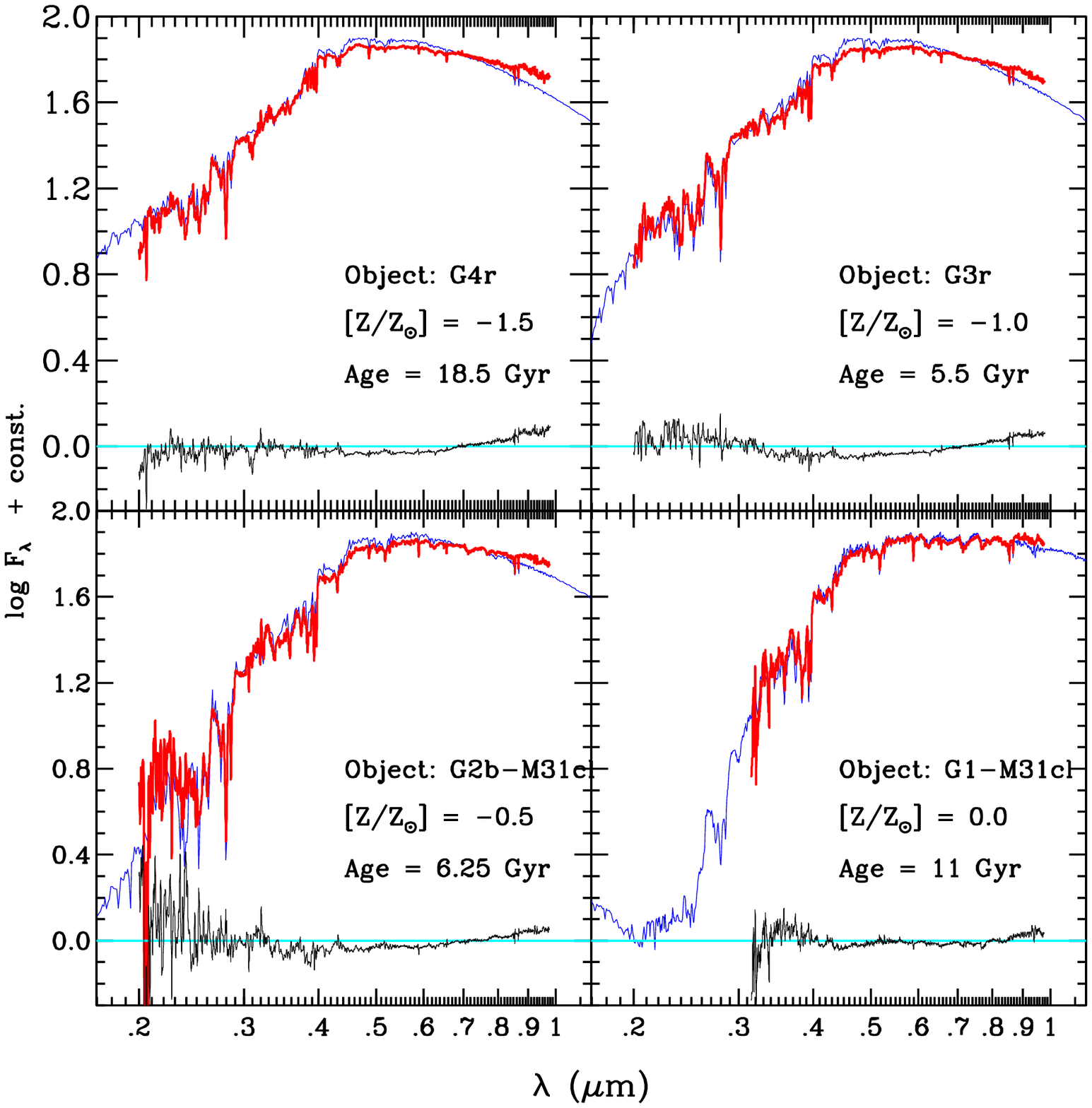}{9cm}{0}{50}{50}{-150}{-80}
\caption{
Same as Fig. 13, but for four different star clusters.
}
\end{figure}

\subsection{Non-solar metallicity}

Figs 13 and 14 show the results of a comparison of SSP models built for
various metallicities using the LCB97 atlas, all for the Salpeter IMF,
with several of the average spectra compiled by Bica et al. (1996a).
The name and the metal content of the observed spectra indicated in
each panel is as given by Bica et al. The quoted age is derived from
the best fit of our model spectra to the corresponding observations.
The residuals (observed - model) are shown in the same vertical
scale.
See the description to Fig. 10 above for more details.
Even though, in detail, the fits for non-solar metallicity stellar
populations are not as good as the ones for solar metallicity, over all
the models reproduce the observations quite well over a wide range of
$[Z/Z_\odot]$, and provide a reliable tool to study these stellar systems. 
The discrepancy can be due both to uncertainties in the synthetic
stellar atlas or the evolutionary tracks at these $[Z/Z_\odot]$.
I have used SSPs in all the fits, neglecting possibly composite stellar
populations, as well as any interstellar reddening.

\begin{figure}
\plotfiddle{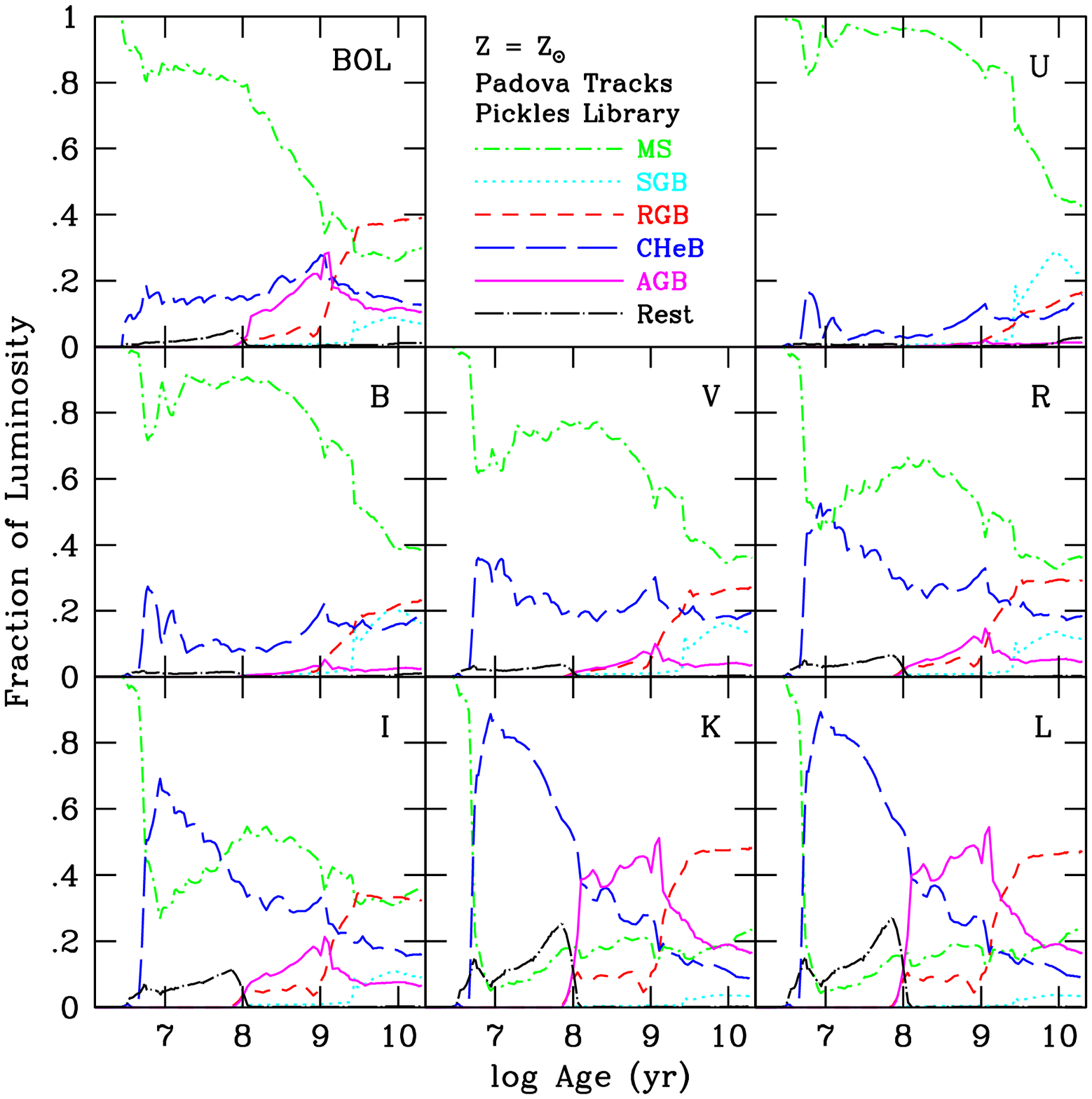}{9cm}{0}{50}{50}{-150}{-80}
\caption{
Contribution of stars in various evolutionary stages
to the bolometric light, and to the broad-band $UBVRIKL$ fluxes for
a $Z = Z_\odot\ $ model SSP computed for the Salpeter IMF
($m_L = 0.1,\ m_U = 125~M_\odot)$ using the $P$ tracks and the
Pickles (1998) stellar atlas.}

\plotfiddle{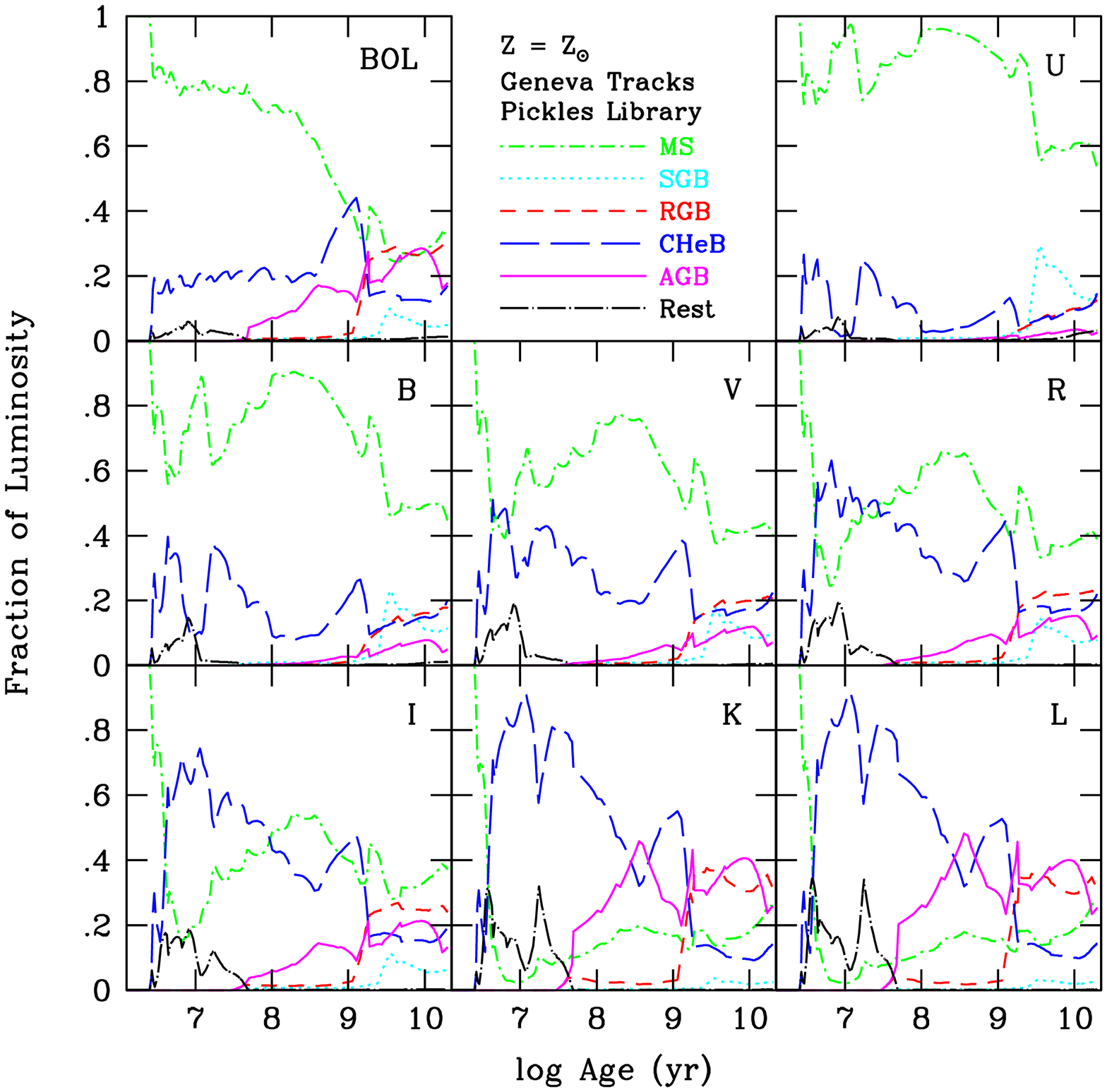}{9cm}{0}{50}{50}{-150}{-80}
\caption{
Contribution of stars in various evolutionary stages
to the bolometric light, and to the broad-band $UBVRIKL$ fluxes for
a $Z = Z_\odot\ $ model SSP computed for the Salpeter IMF
($m_L = 0.1,\ m_U = 125~M_\odot)$ using the $G$ tracks and the
Pickles (1998) stellar atlas.} 
\end{figure}

\section{Different sources of uncertainties in population synthesis models}

\subsection{Uncertainties in the astrophysics of stellar evolution}

There are significant differences in the fractional contribution
to the integrated light by red giant branch (RGB) and asymptotic giant branch
(AGB) stars in SSPs computed for different sets of evolutionary tracks.
Fig. 15 shows the contribution of stars in various evolutionary stages
to the bolometric light, and
to the broad-band $UBVRIKL$ fluxes for a $Z = Z_\odot\ $ model SSP computed
for the Salpeter IMF
($m_L = 0.1,\ m_U = 125~M_\odot)$ using the $P$ tracks and the
Pickles (1998) stellar atlas. 
The meaning of each line is indicated in the top central frame.
Fig. 16 shows the corresponding plot
for an equivalent model computed according to the $G$ tracks.
The contribution of the RGB stars is higher in the $P$ track model than in
the $G$ track model.
Correspondingly, the AGB stars contribute less in the $P$ track model than in
the $G$ track model.
For instance, for $t > 1$ Gyr, RGB and AGB stars contribute 40\% and 10\%,
respectively, to the bolometric light in the $P$ track model (Fig. 15).
These fractions change to 30\% and 20\% in the $G$ track model (Fig. 16).
These differences are seeing more clearly in Fig. 17 which shows the
ratio of the fractional contribution by different stellar groups
in the $G$ track model to that in the $P$ track model.
According to the fuel consumption theorem (Renzini 1981), these numbers reflect
relatively large differences in the amount of fuel used up in the RGB and AGB
phases by stars of the same mass and initial chemical composition depending on
the stellar evolutionary code.


Fig. 20 shows the difference in $B$ magnitude and $B-V$ and
$V-K$ color between a $G$ track model SSP and a $P$ track model SSP as
seen both in the rest frame of the galaxy (vs. galaxy age in the left hand
side panels) and the observer frame (vs. redshift in the right hand side
panels). These differences reach quite substantial values.
The observer frame quantities include both the $k$ and the evolutionary
corrections. Here and elsewhere in this paper
I assume $H_0 = 65$ km s$^{-1}$ Mpc$^{-1}$, $\Omega = 0.10$, and the age
of galaxies to be $tg = 12$ Gyr.

\subsection{On the energetics of model stellar populations}

Buzzoni (1999) has argued that most population synthesis models violate
basic prescriptions from the fuel consumption theorem (FCT). Fig. 18
should be compared to Fig. 2 of Buzzoni (1999). The line with square dots
along it is reproduced from Buzzoni's Fig. 2. The other lines show the
dependence of the ratio of the Post-MS to MS contribution to the bolometric
flux for different models. The heavy lines correspond to the Salpeter IMF
models. The thin lines to the Scalo IMF models. The solid lines correspond to
$P$ track models, whereas the dashed lines correspond to the $G$ track
models. The $G$ track model for the Salpeter IMF (heavy dashed line) is
in quite good agreement with Buzzoni's model for $t > 5$ Gyr.

Fig. 19 (after Buzzoni's Fig. 1) plots the $M/L_V$ ratio vs. the Post-MS to
MS contribution in the $V$ band. The open dots correspond to the
models shown in Buzzoni's Fig. 1. The solid dot is Buzzoni's model marked
B in his Fig. 1. The solid triangles correspond to our $Z_\odot$ SSP models
for various stellar atlas using the $P$ tracks and the Salpeter IMF.
The open triangles are for the same models but for the Scalo IMF.
The solid pentagons represent the $G$ track models for the Salpeter
IMF and the open pentagons the same models but for the Scalo IMF.
The three solid squares joined by a line represent sub-solar metallicity
models for the $P$ tracks and the Salpeter IMF. The three open squares
joined by a line are for identical models using the Scalo IMF.
Fig. 19 shows clearly that the position of points representing various models
in this diagram is a strong function of the stellar IMF, the set of evolutionary
tracks, and the chemical composition of the stellar population.
It may be too simplistic to attribute the dispersion of the points to
a violation of the FCT (Buzzoni 1999).

\begin{figure}
\plotfiddle{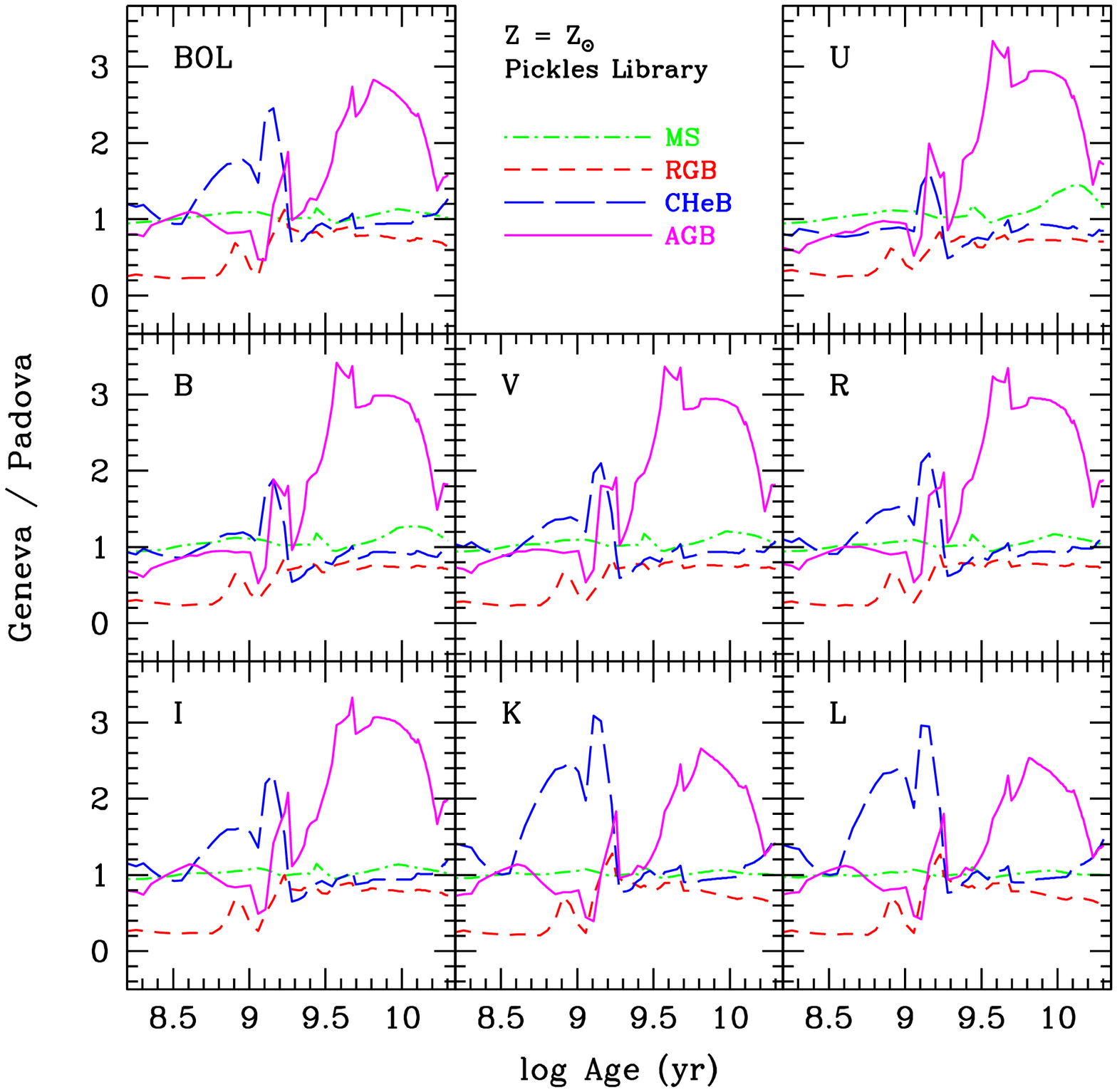}{9cm}{0}{50}{50}{-150}{-80}
\caption{
Ratio of the fractional contribution by different stellar groups
in the $G$ track model to that in the $P$ track model.}

\plotfiddle{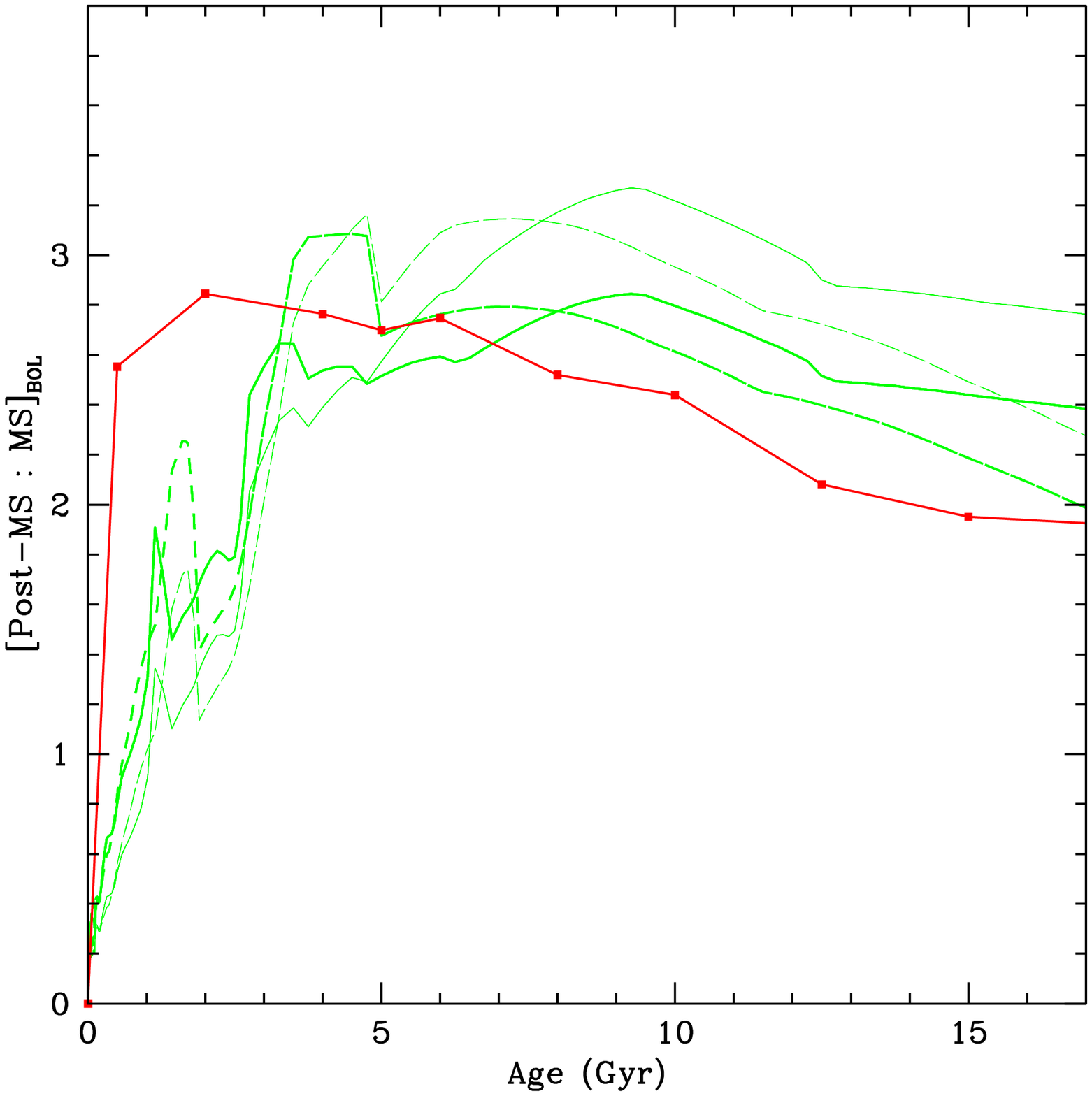}{9cm}{0}{50}{50}{-150}{-80}
\caption{
Ratio of the Post-MS to MS contribution to the bolometric
flux vs. age for different models.
This figure should be compared to Fig. 2 of Buzzoni (1999).
The line with square dots along it is reproduced from Buzzoni's Fig. 2.
The heavy lines correspond to the Salpeter IMF models.
The thin lines to the Scalo IMF models. The solid lines correspond to
$P$ track models, whereas the dashed lines correspond to the $G$ track
models. The $G$ track model for the Salpeter IMF (heavy dashed line) is
in quite good agreement with Buzzoni's model for $t > 5$ Gyr.
}
\end{figure}

\begin{figure}
\plotfiddle{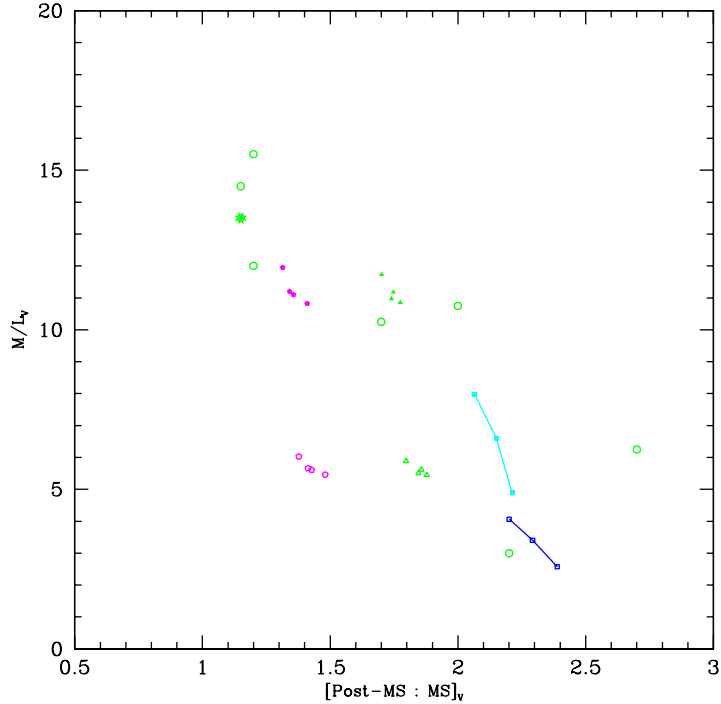}{10cm}{0}{50}{50}{-150}{-80}
\caption{
$M/L_V$ ratio vs. the Post-MS to MS contribution in the $V$ band.
This figure should be compared to Fig. 1 of Buzzoni (1999).
The open dots correspond to the models shown in Buzzoni's Fig. 1.
The solid dot is Buzzoni's model marked B in his Fig. 1.
The solid triangles correspond to our $Z_\odot$ SSP models
for various stellar atlas using the $P$ tracks and the Salpeter IMF.
The open triangles are for the same models but for the Scalo IMF.
The solid pentagons represent the $G$ track models for the Salpeter
IMF and the open pentagons the same models but for the Scalo IMF.
The three solid squares joined by a line represent sub-solar metallicity
models for the $P$ tracks and the Salpeter IMF. The three open squares
joined by a line are for identical models using the Scalo IMF.
}
\end{figure}

\begin{figure}
\plotfiddle{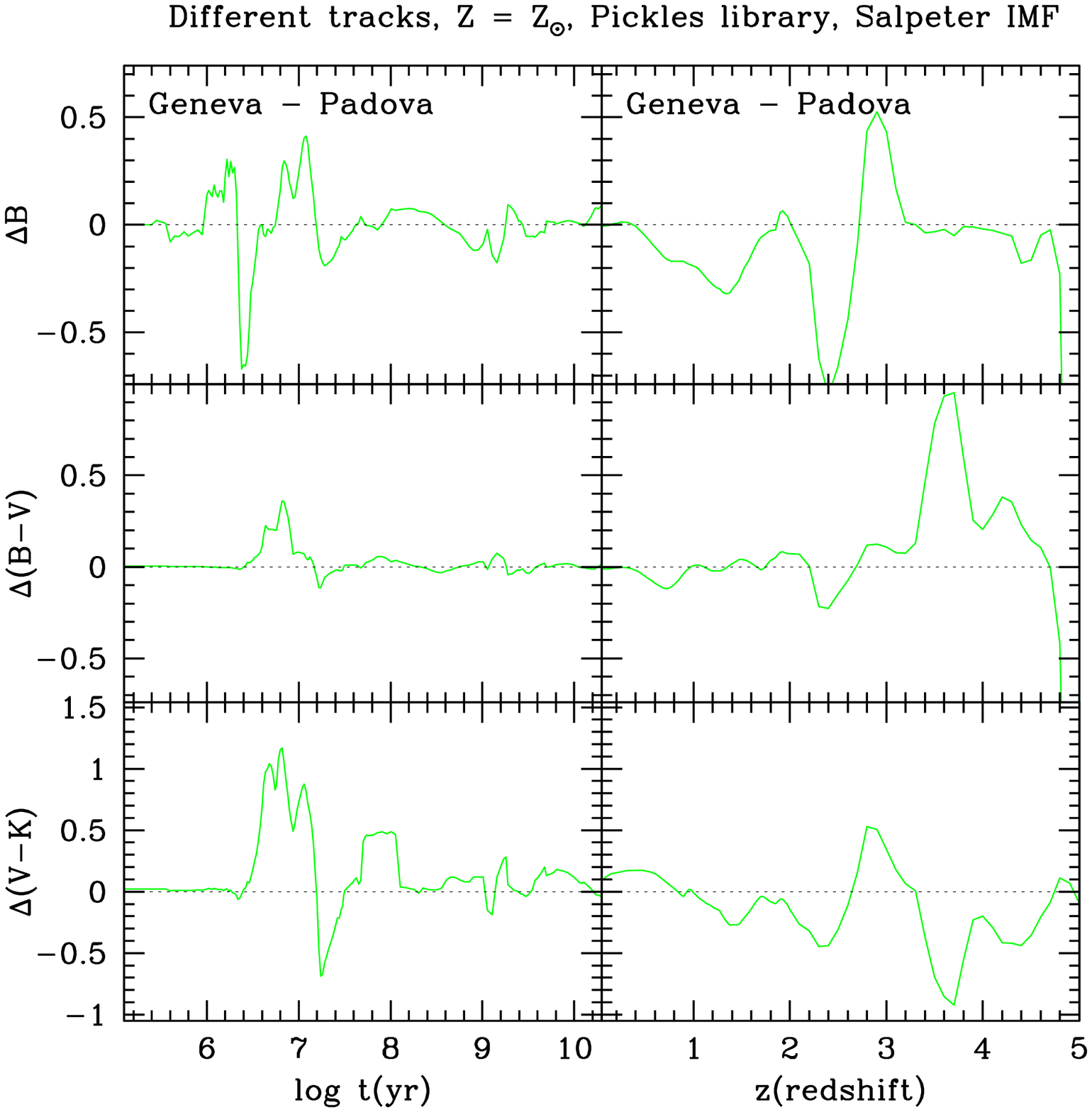}{10cm}{0}{50}{50}{-150}{-80}
\caption{
Difference in $B$ magnitude and $B-V$ and $V-K$ color
between a $G$ track model SSP and a $P$ track model SSP,
vs. age in the galaxy rest frame (LHS panels)
and vs. redshift in the observer frame (RHS panels).
}

\plotfiddle{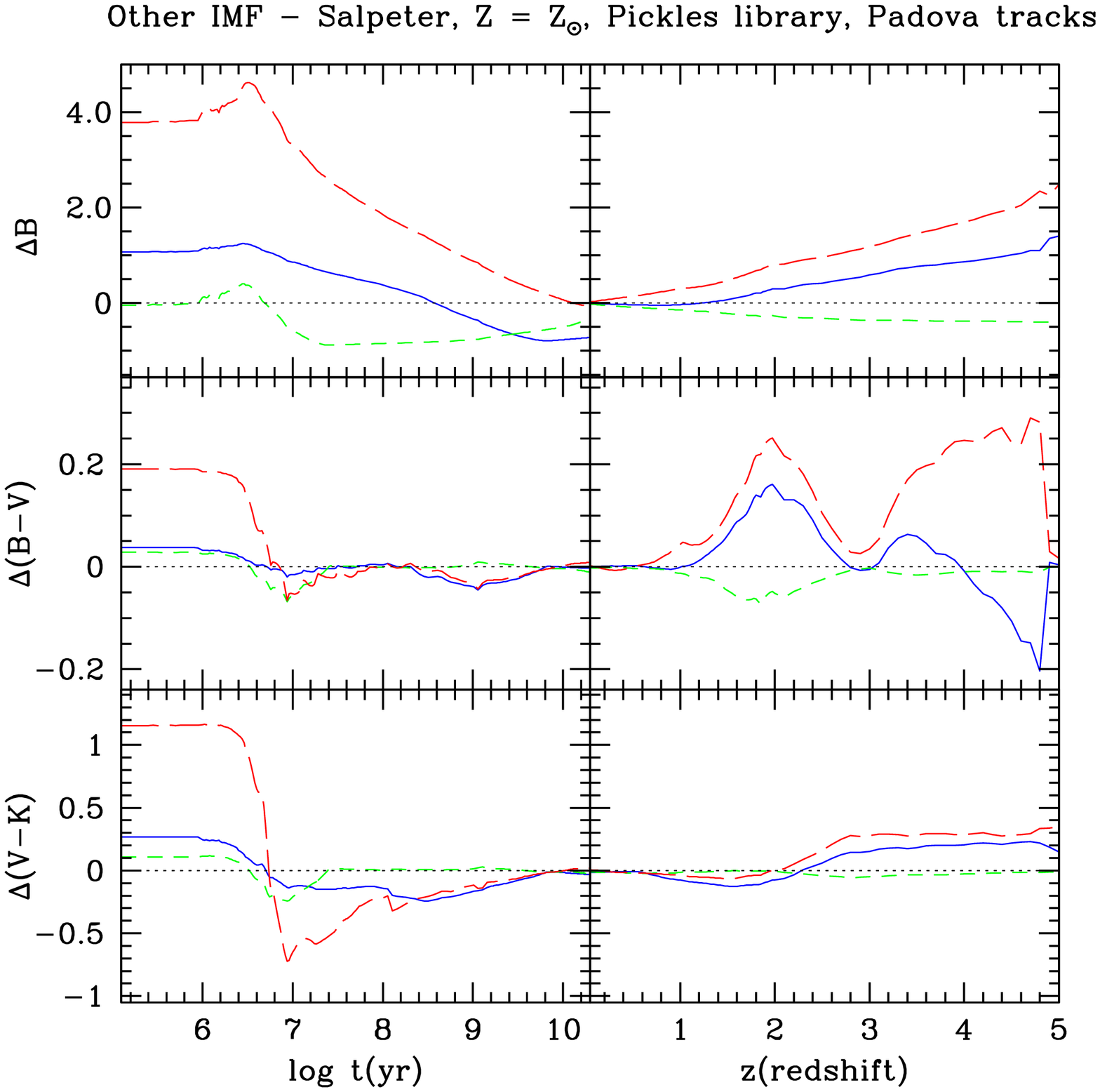}{10cm}{0}{50}{50}{-150}{-80}
\caption{
Difference in $B$ magnitude and $B-V$ and $V-K$ color 
for various IMF, $P$ track model SSP's,
with respect to the Salpeter IMF model,
vs. galaxy age in the galaxy rest frame (LHS panels)
and vs. redshift in the observer frame (RHS panels).
}
\end{figure}

\begin{figure}
\plotfiddle{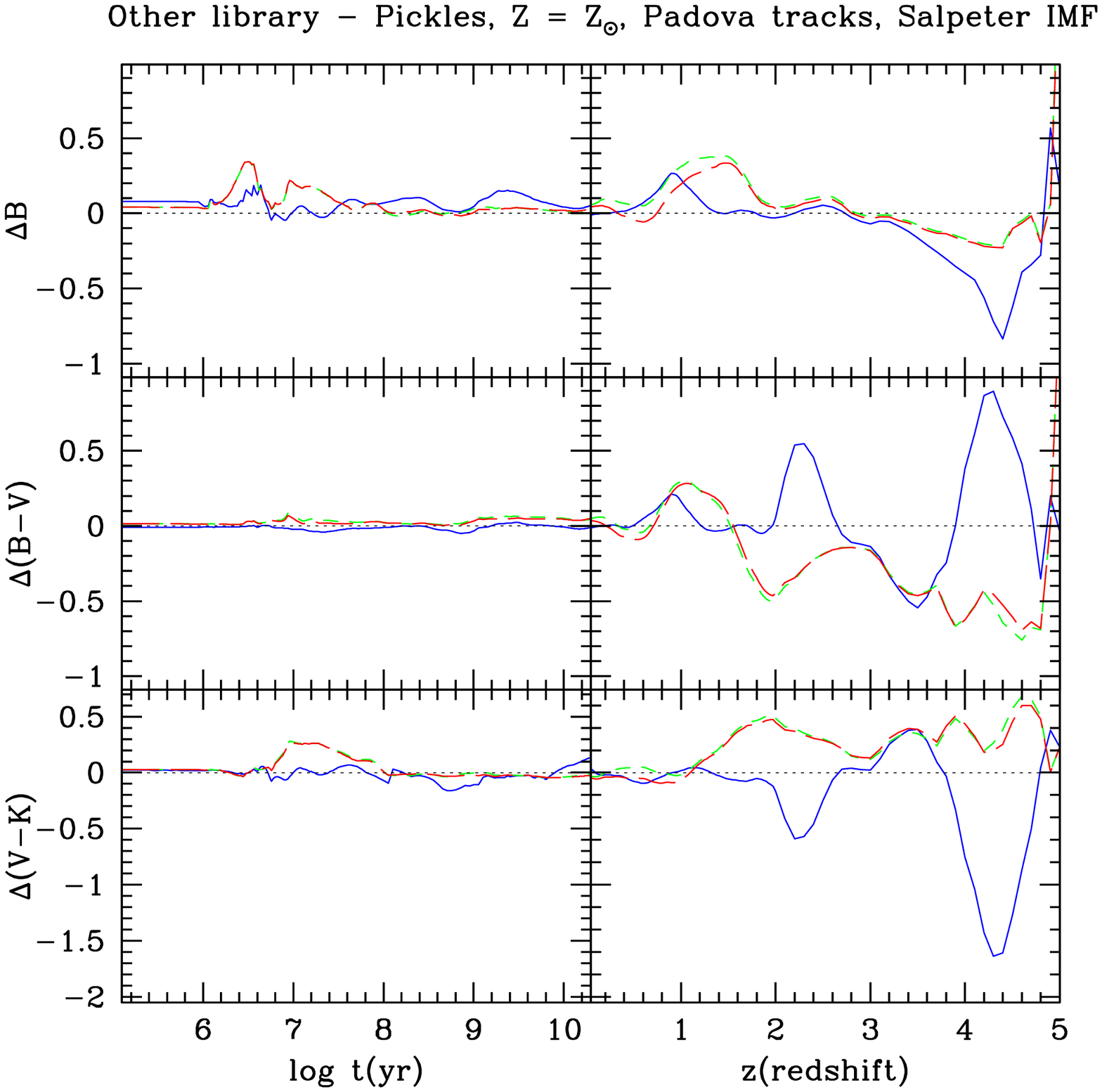}{10cm}{0}{50}{50}{-150}{-80}
\caption{
Difference in $B$ magnitude and $B-V$ and $V-K$ color for $P$ track model
SSP's computed for various $Z=Z_\odot$ stellar libraries,
with respect to the SSP model computed with the Pickles (1998) stellar atlas,
vs. galaxy age in the galaxy rest frame (LHS panels)
and vs. redshift in the observer frame (RHS panels).
}

\plotfiddle{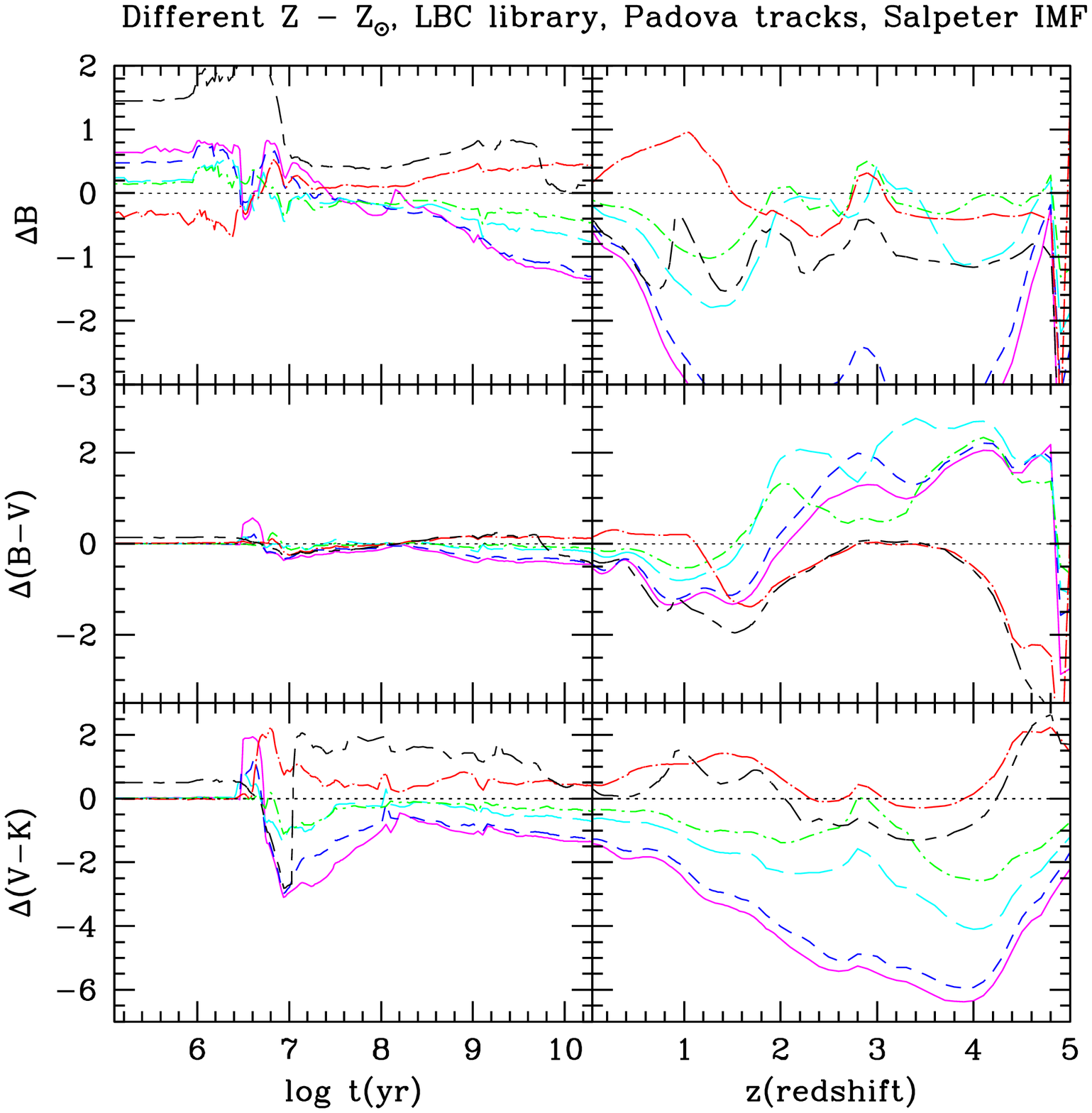}{10cm}{0}{50}{50}{-150}{-80}
\caption{
Differences between SSP models for non-solar
composition and the solar case,
vs. galaxy age in the galaxy rest frame (LHS panels)
and vs. redshift in the observer frame (RHS panels).
}
\end{figure}

\subsection{Uncertainties in the stellar IMF}

It is constructive to compare models computed for identical ingredients
except for the stellar IMF. Fig. 21 shows the results of such a comparison.
Brightness and color differences with respect to the Salpeter IMF model
SSP are shown vs. galaxy age in the galaxy rest frame (LHS panels) and vs.
redshift in the observer frame (RHS panels) for SSP $P$ track models
computed for the following IMFs:
Scalo (1986, {\it solid line}),
Miller \& Scalo (1979, {\it short dashed line}), and
Kroupa et al. (1993, {\it long dashed line}).

Fig. 22 compares in the same format as before the results of using different
solar metallicity stellar libraries for a $P$ track SSP model.
Brightness and color differences with respect to the SSP model computed
with the Pickles (1998) stellar atlas are shown vs. galaxy age in the galaxy
rest frame (LHS panels) and vs.
redshift in the observer frame (RHS panels) for SSP $P$ track models
computed for the following stellar libraries:
Extended Gunn \& Stryker atlas used by BC93 ({\it solid line}),
LCB97 uncorrected atlas ({\it short dashed line}), and
LCB97 corrected atlas ({\it long dashed line}).

\subsection{Different chemical composition}

In this section we explore the differences between SSP models for non-solar
composition and the solar case. Fig. 23 shows the brightness and color
differences with respect to the SSP model for $Z = Z_\odot$. All models shown
in this figure are for the $P$ tracks and use the LCB97 stellar atlas.
The lines in this figure have the following meaning:
$Z = 0.0001$ ({\it solid line}),
$Z = 0.0004$ ({\it short dashed line}),
$Z = 0.004$ ({\it long dashed line}),
$Z = 0.008$ ({\it dot - short dashed line}),
$Z = 0.05$ ({\it short dash - long dashed line}), and
$Z = 0.1$ ({\it dot - long dashed line}).

\subsection{Different history of chemical evolution}

Fig. 24 shows three possible chemical evolutionary histories, $Z(t)$,
quick ($Z_Q$, {\it short dashed line}),
linear ($Z_L$, {\it solid line}), and
slow ($Z_S$, {\it long dashed line}),
that reach $Z=0.1=5 \times Z_\odot$ at 15 Gyr.
The dotted lines indicate $Z_\odot$ and $t_g = 12$ Gyr.
I have computed models for a SFR $\Psi(t) \propto exp(-t/\tau)$,
with $\tau = 5$ Gyr, which evolve chemically accordingly to
the lines shown in Fig. 24. The difference in brightness and
color of these three models
with respect to a $Z=Z_\odot$ model for the same SFR are shown in
Fig. 25. The meaning of the lines is as follows:
$Z(t) = Z_Q$ ({\it short dashed line}),
$Z(t) = Z_L$ ({\it solid line}), and
$Z(t) = Z_S$ ({\it long dashed line}).

\begin{figure}
\plotfiddle{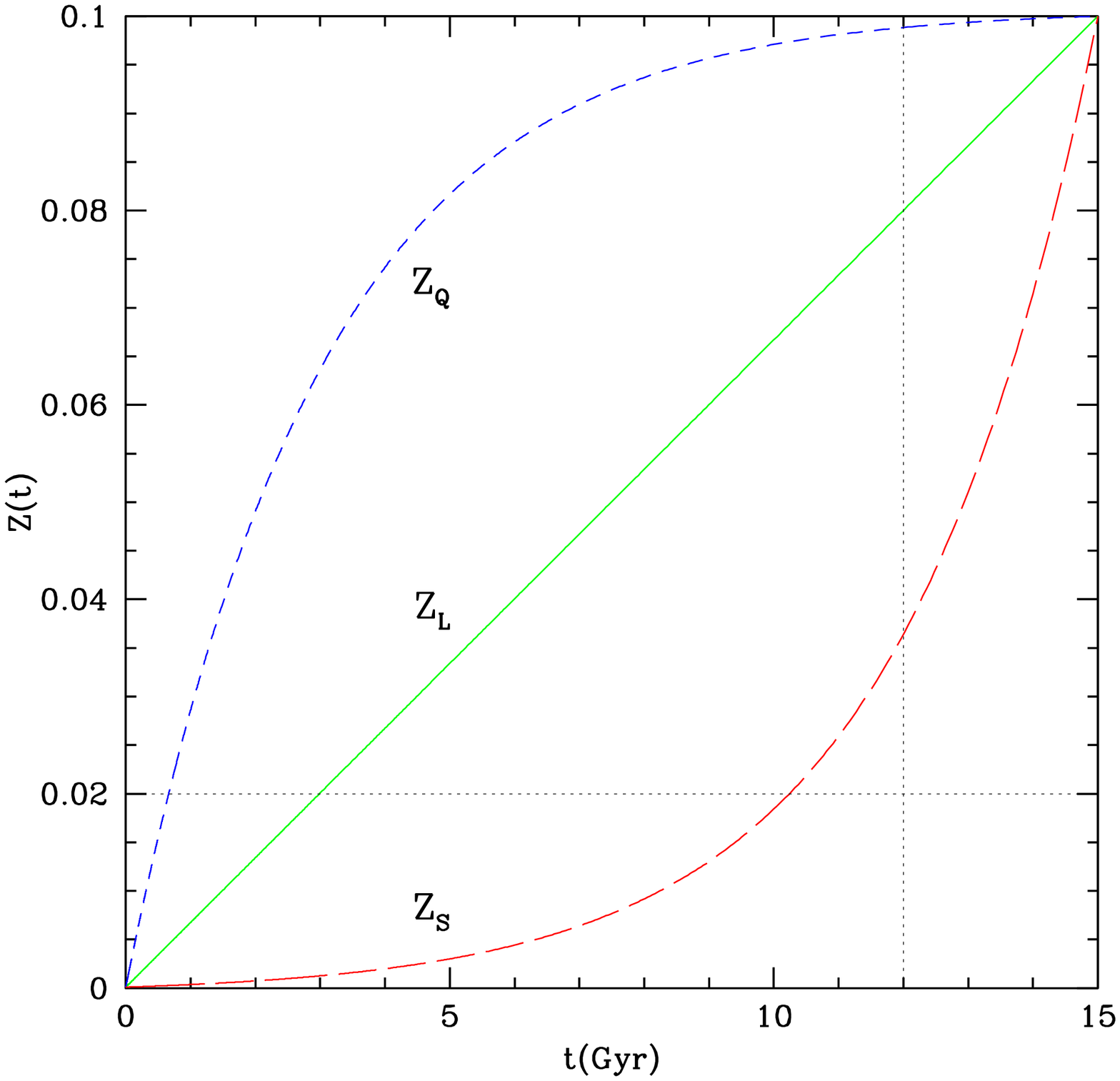}{10cm}{0}{50}{50}{-150}{-80}
\caption{
Three possible chemical evolutionary histories, $Z(t)$,
quick ($Z_Q$, {\it short dashed line}),
linear ($Z_L$, {\it solid line}), and
slow ($Z_S$, {\it long dashed line}),
that reach $Z=0.1=5 \times Z_\odot$ at 15 Gyr.
}

\plotfiddle{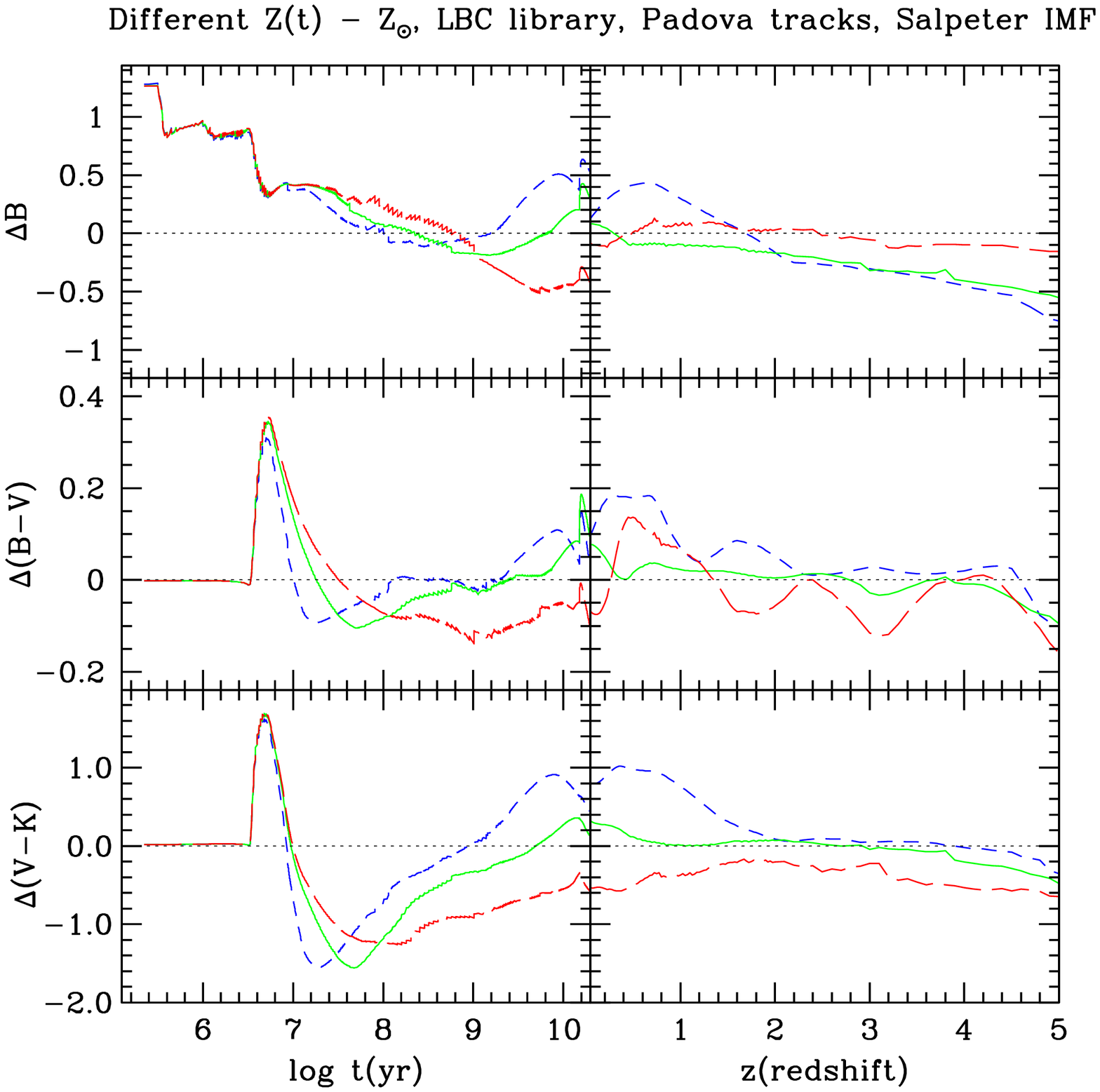}{10cm}{0}{50}{50}{-150}{-80}
\caption{
Difference in brightness and color of three models
that evolve chemically according to Fig. 24
with respect to a $Z=Z_\odot$ model for the same SFR,
vs. galaxy age in the galaxy rest frame (LHS panels)
and vs. redshift in the observer frame (RHS panels).
}
\end{figure}

\subsection{Evolution in the observer frame at various cosmological epochs}

In Figs. 26 to 31, I summarize the range of values expected in the measured
$(V-R)$ and $(V-K)$ colors in the observer frame at various redshifts $z$
as a function of galaxy age.
In these figures,
the panel marked $TRACKS$ shows the range of colors obtained for solar
metallicity SSP models computed using the Pickles empirical stellar atlas
with the Salpeter IMF for the $P$ and the $G$ tracks.
In the panel marked $IMF$ I show $Z=Z_\odot$ SSP models computed for the $P$
tracks, the Pickles stellar atlas, and the Salpeter, the Scalo, and the
Miller-Scalo IMFs.
The panel marked $SEDs$ shows the evolution of $Z=Z_\odot$ SSP models computed
with the $P$ tracks and the Salpeter IMF, using the empirical Gunn-Stryker
and Pickles stellar libraries, as well as the original and corrected versions
of the LCB atlas for $Z=Z_\odot$.
The panel marked $SFR$ shows the evolution of an SSP model together with a model
in which stars form at a constant rate during the first Gyr in the life of
the galaxy (1 Gyr model),
both computed with the $P$ tracks, Salpeter IMF, and the
Pickles stellar library.
The panel marked $Z$ shows the range of colors covered by SSP models of
metallicity $Z = 0.004, 0.008$ and 0.02 (solar), computed with the $P$ tracks
and the Salpeter IMF. In the solar case, I repeat the models shown in the
panel marked $SEDs$.
The panel marked $ALL$ summarizes the results of the previous panels.
The reddest color obtained at any age in the previous 5 panels is shown
as the top solid line. The bluest color is shown as a dotted line. The
average color is indicated by the solid line between these two extremes.
The 1 Gyr model is shown as a dashed line to show the dominant effects of star
formation in galaxy colors.

Figs. 26 and 27 show the range of values expected in $(V-R)$ and $(V-K)$
in the observer frame at $z=0$ as a function of galaxy age.
The maximum age allowed is the age of the universe, $t_u = 13.5$ Gyr
at $z=0$ using $H_0 = 65$ km s$^{-1}$ Mpc$^{-1}$, $\Omega = 0.10$.
Figs. 28 and 29 show the same quantities but for galaxies seen at $z=1.552$.
The age of the universe for this $z$ in this cosmology is $t_u=4.6$ Gyr.
Figs. 30 and 31 correspond to $z=3$, in this case $t_u=2.7$ Gyr.
From Figs. 26 to 31 I conclude that metallicity $Z$ and the SFR are the
most dominant factors determining the range of allowed colors.

The horizontal lines shown across the panels in Figs. 28 and 29 indicate
the color $\pm \sigma$ of the galaxy LBDS 53W091 observed by Spinrad et al.
(1997). Our models reproduce the colors of this galaxy at an age close to
1.5 Gyr.

Figs. 32 to 35 are based on the panel marked $ALL$ of Figs. 26 to 29, and
similar figures for $(V-U)$, $(V-B)$, $(V-I)$, and $(V-J)$ not shown in
this work. To build these figures I have subtracted from each line in the
previous figures, the
color of the $Z=Z_\odot$ SSP model computed with the $P$ tracks, the
Salpeter IMF, and the Pickles stellar library.
We conclude that the evolution of $(V-R)$ is less model dependent than for
any other color shown in these figures.

Figs. 36 to 39 are also based on the panel marked $ALL$ of Figs. 26 to 29
and similar figures for other values of $z$ not shown in this work.
Figs. 36 to 39 show the evolution in time of $(V-U)$, $(V-B)$, $(V-R)$, and
$(V-K)$ in the observer frame for several values of the redshift $z$.
The color of the $Z=Z_\odot$ SSP model computed with the $P$ tracks, the
Salpeter IMF, and the Pickles stellar library has been subtracted from
the lines in the previous figures.
Again, $(V-R)$ shows less variations with model than the other colors.

\begin{figure}
\plotfiddle{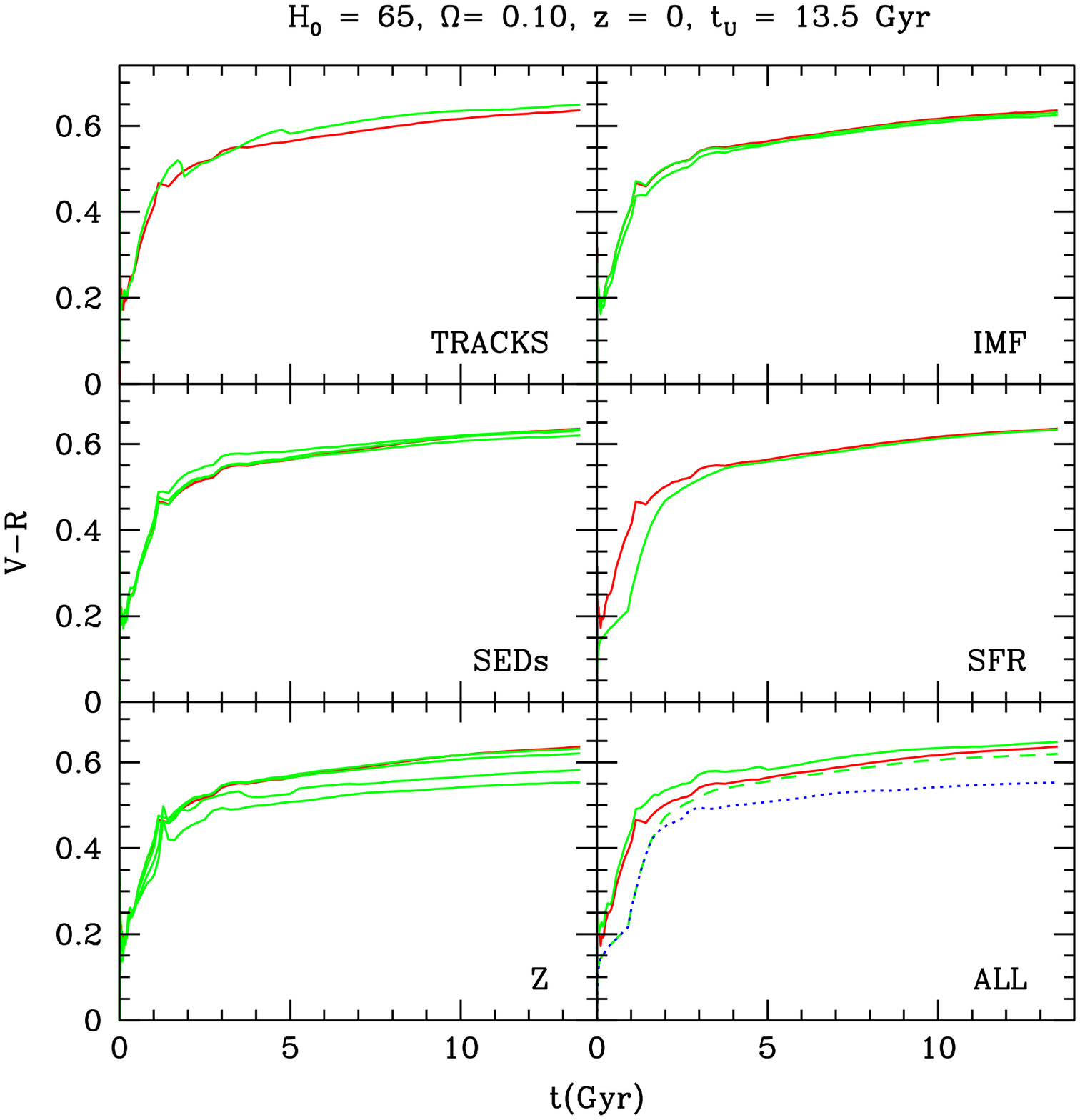}{10cm}{0}{50}{50}{-150}{-80}
\caption{
$(V-R)$ vs. time in the observer frame at $z=0$. See \S11.6 for details.
}
\plotfiddle{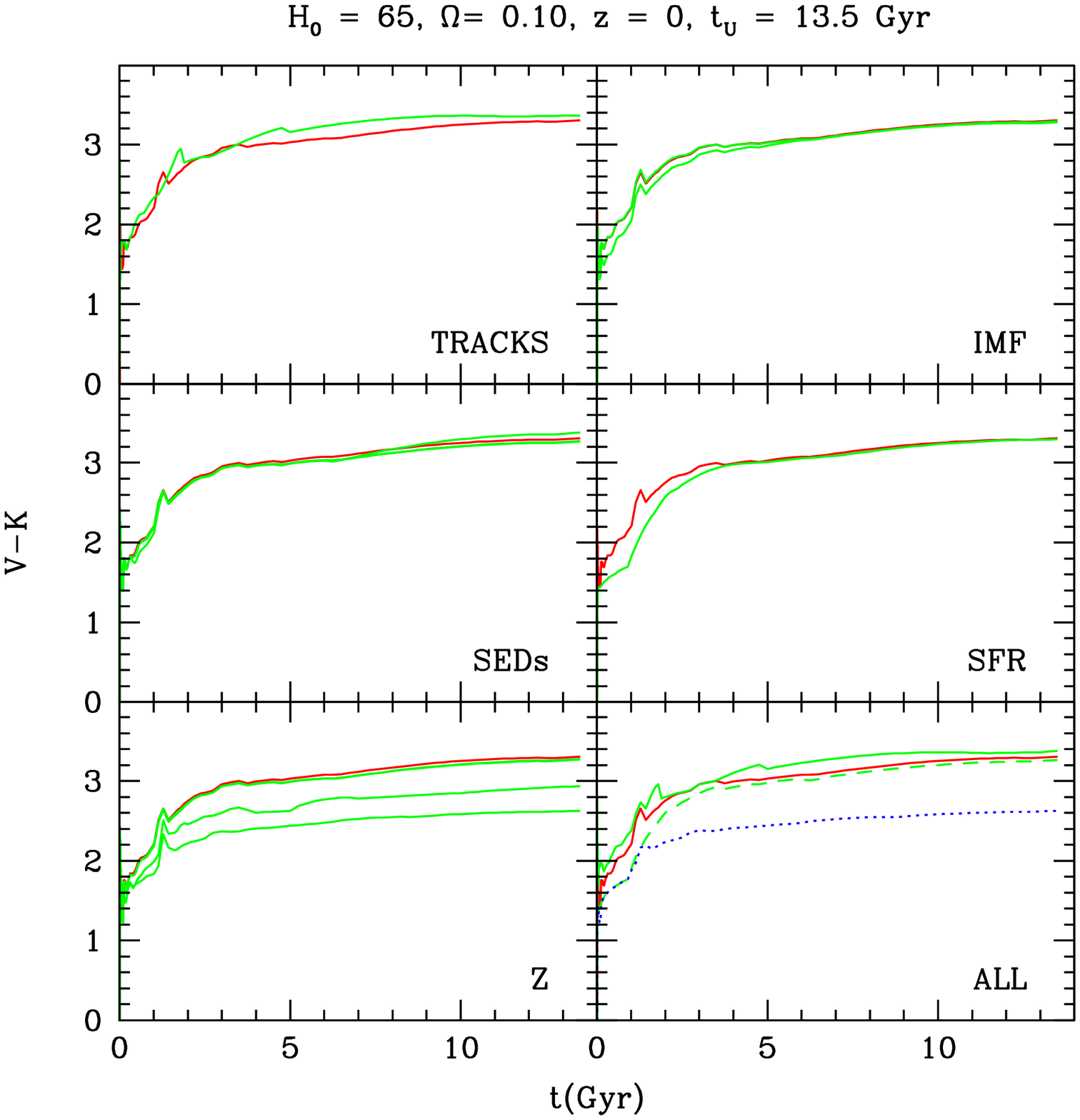}{10cm}{0}{50}{50}{-150}{-80}
\caption{
$(V-K)$ vs. time in the observer frame at $z=0$. See \S11.6 for details.
}
\end{figure}

\begin{figure}
\plotfiddle{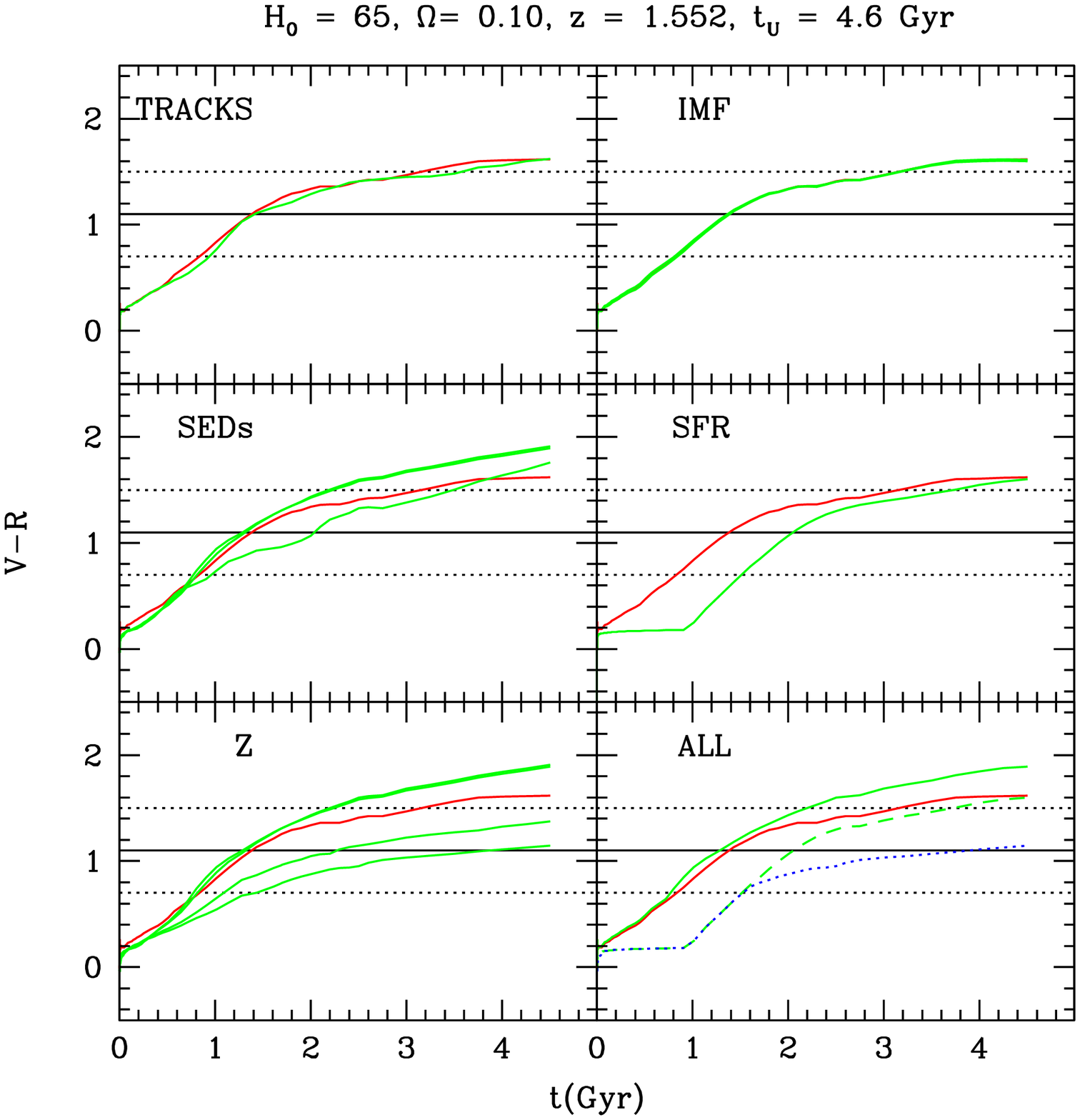}{10cm}{0}{50}{50}{-150}{-80}
\caption{
$(V-R)$ vs. time in the observer frame at $z=1.552$.
See \S11.6 for details.
}
\plotfiddle{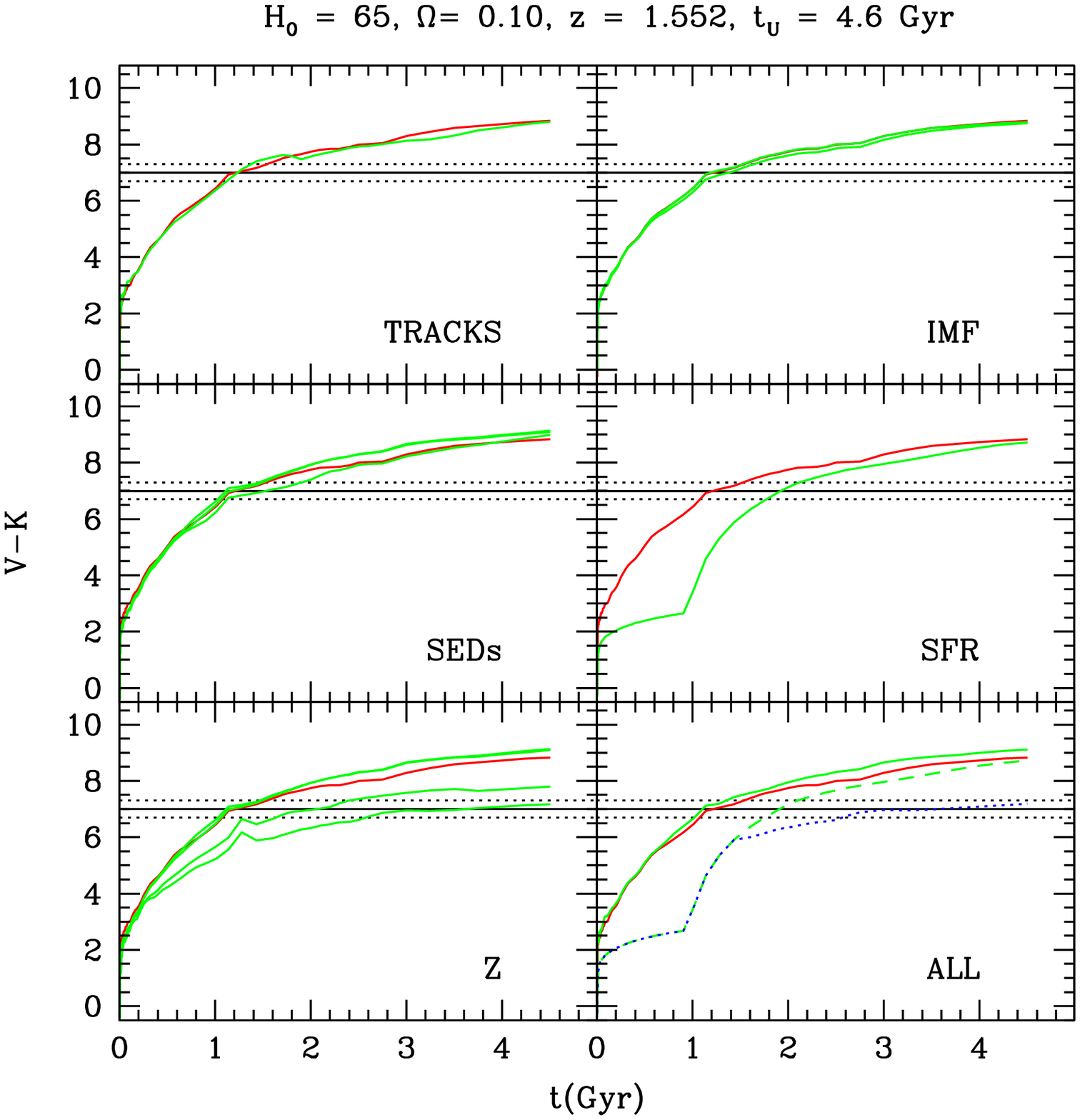}{10cm}{0}{50}{50}{-150}{-80}
\caption{
$(V-K)$ vs. time in the observer frame at $z=1.552$.
See \S11.6 for details.
}
\end{figure}

\begin{figure}
\plotfiddle{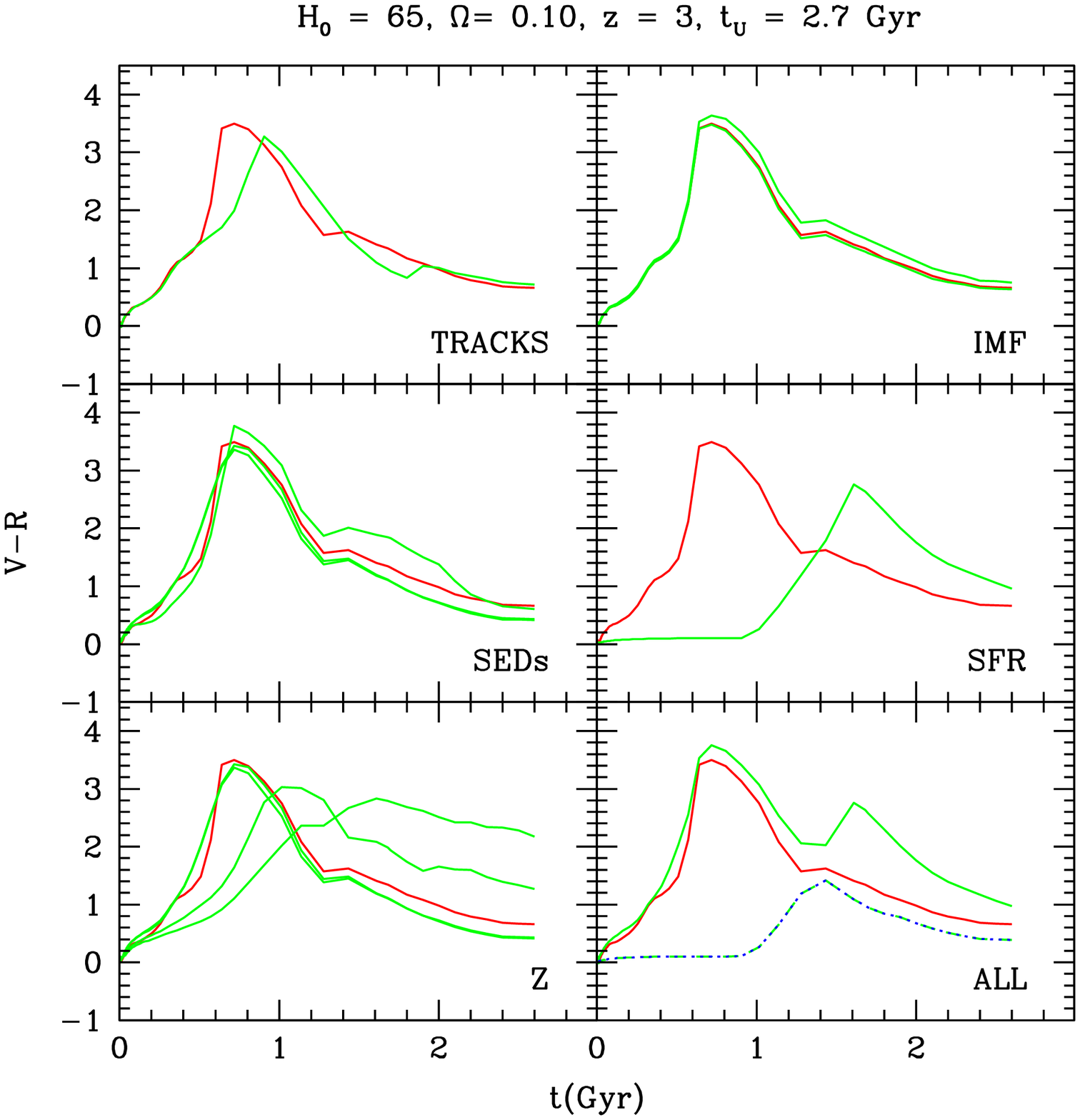}{10cm}{0}{50}{50}{-150}{-80}
\caption{
$(V-R)$ vs. time in the observer frame at $z=3$. See \S11.6 for details.
}
\plotfiddle{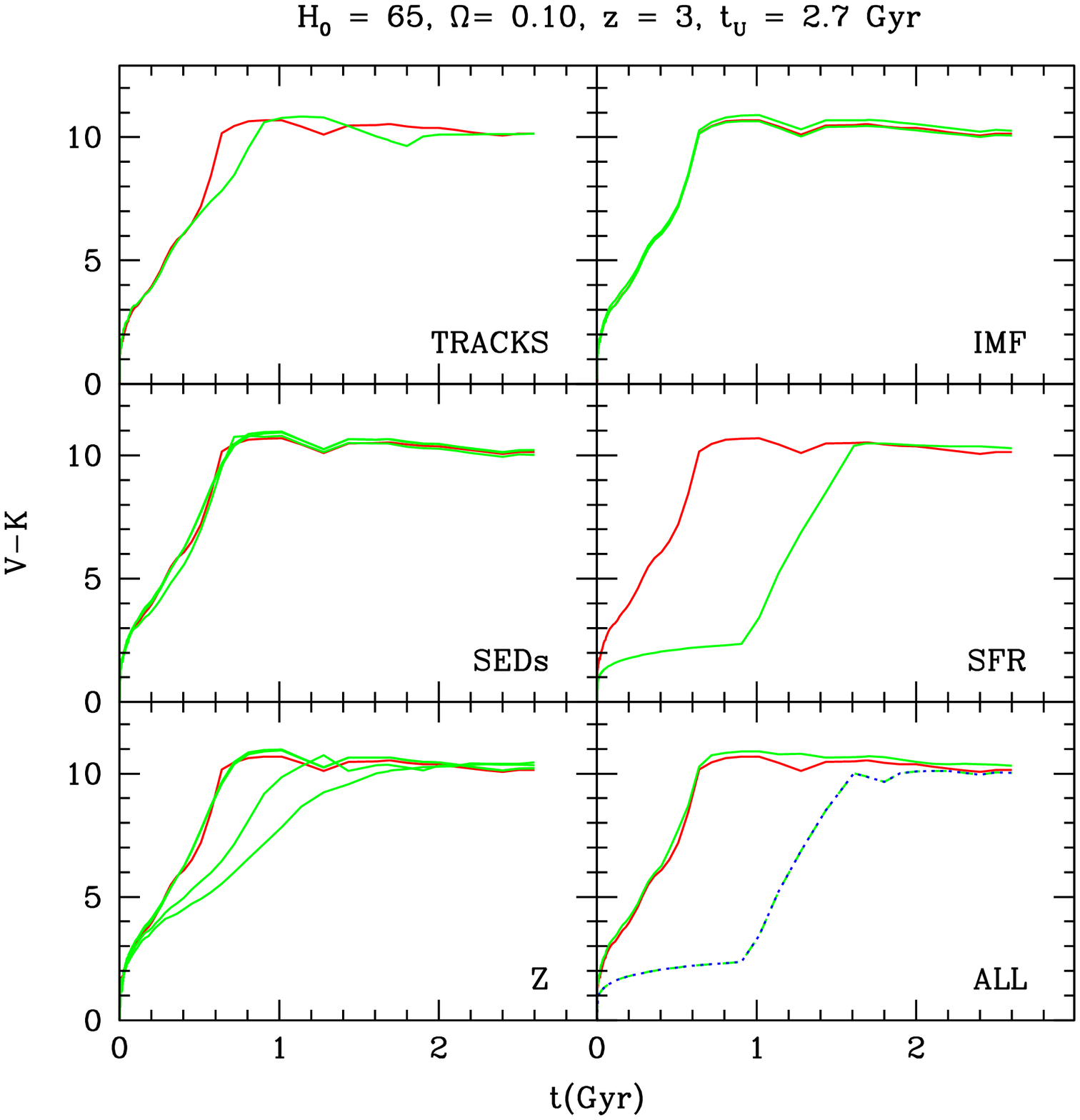}{10cm}{0}{50}{50}{-150}{-80}
\caption{
$(V-K)$ vs. time in the observer frame at $z=3$. See \S11.6 for details.
}
\end{figure}

\begin{figure}
\plotfiddle{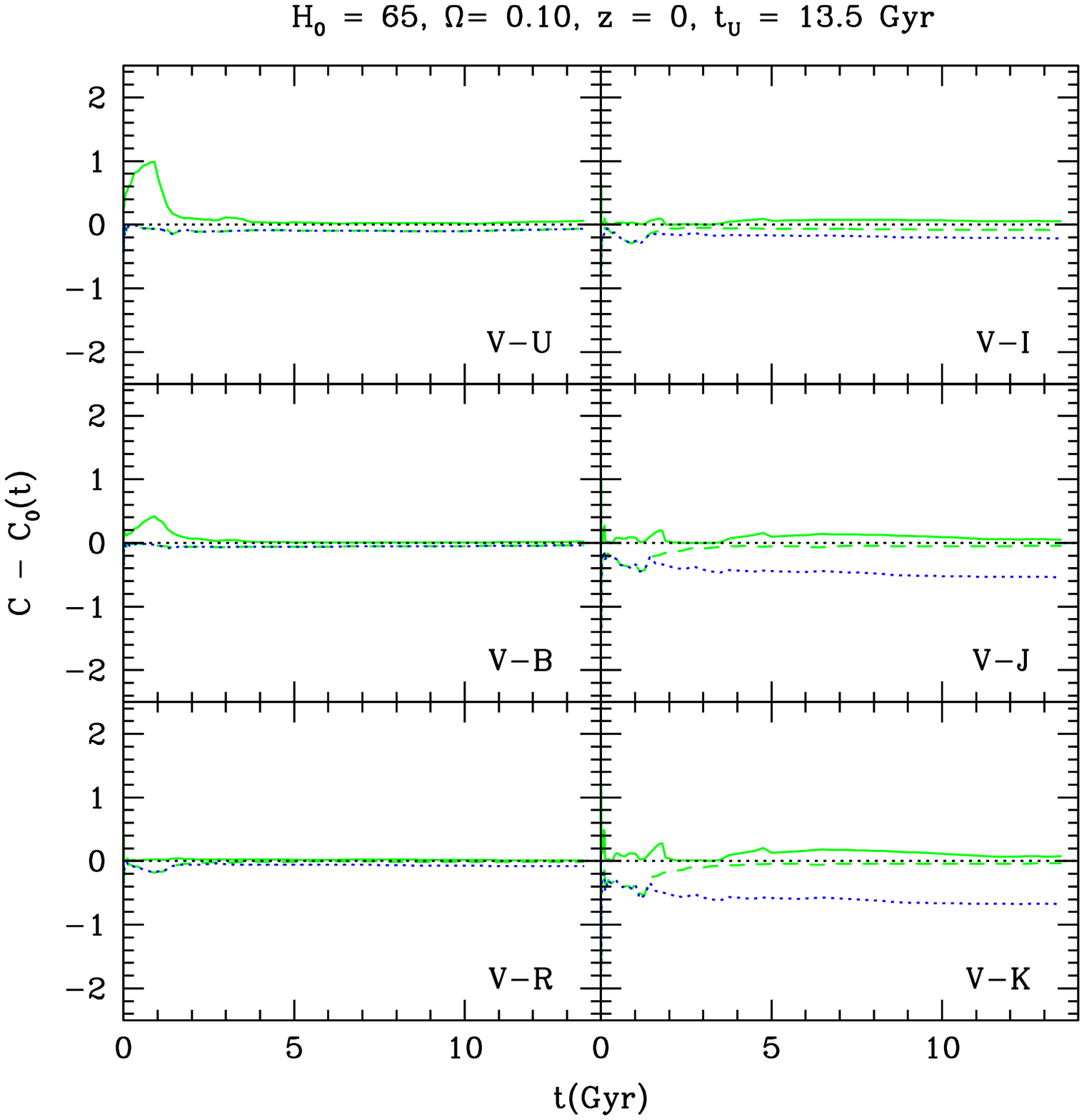}{10cm}{0}{50}{50}{-150}{-80}
\caption{
Color vs. time in the observer frame at $z=0$.
The color of the $Z=Z_\odot$ SSP model computed with the $P$ tracks, the
Salpeter IMF, and the Pickles stellar library have been subtracted from
each line. See \S11.6 for details.
}
\plotfiddle{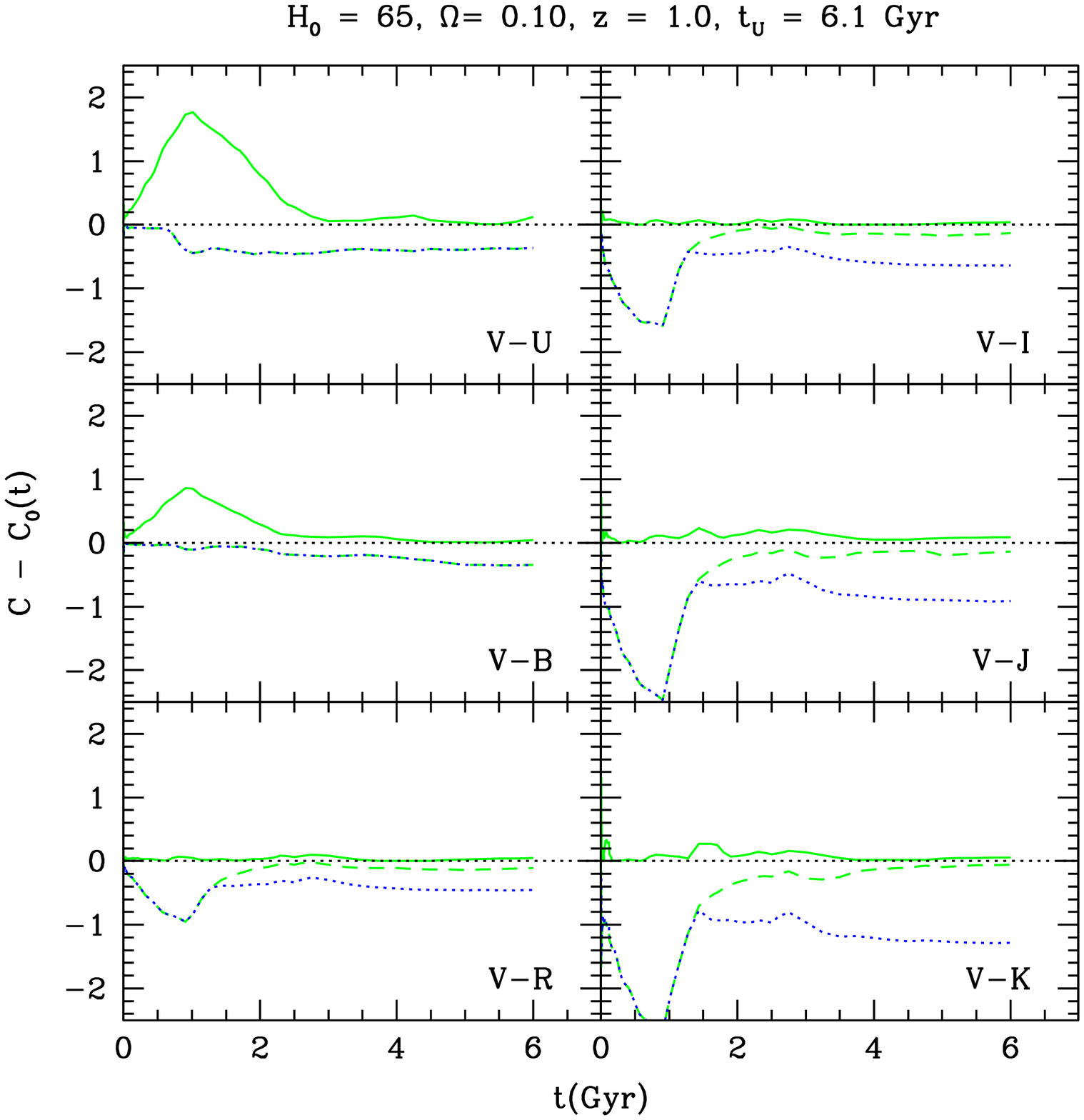}{10cm}{0}{50}{50}{-150}{-80}
\caption{
Color vs. time in the observer frame at $z=1$.
The color of the $Z=Z_\odot$ SSP model computed with the $P$ tracks, the
Salpeter IMF, and the Pickles stellar library have been subtracted from
each line. See \S11.6 for details.
}
\end{figure}

\begin{figure}
\plotfiddle{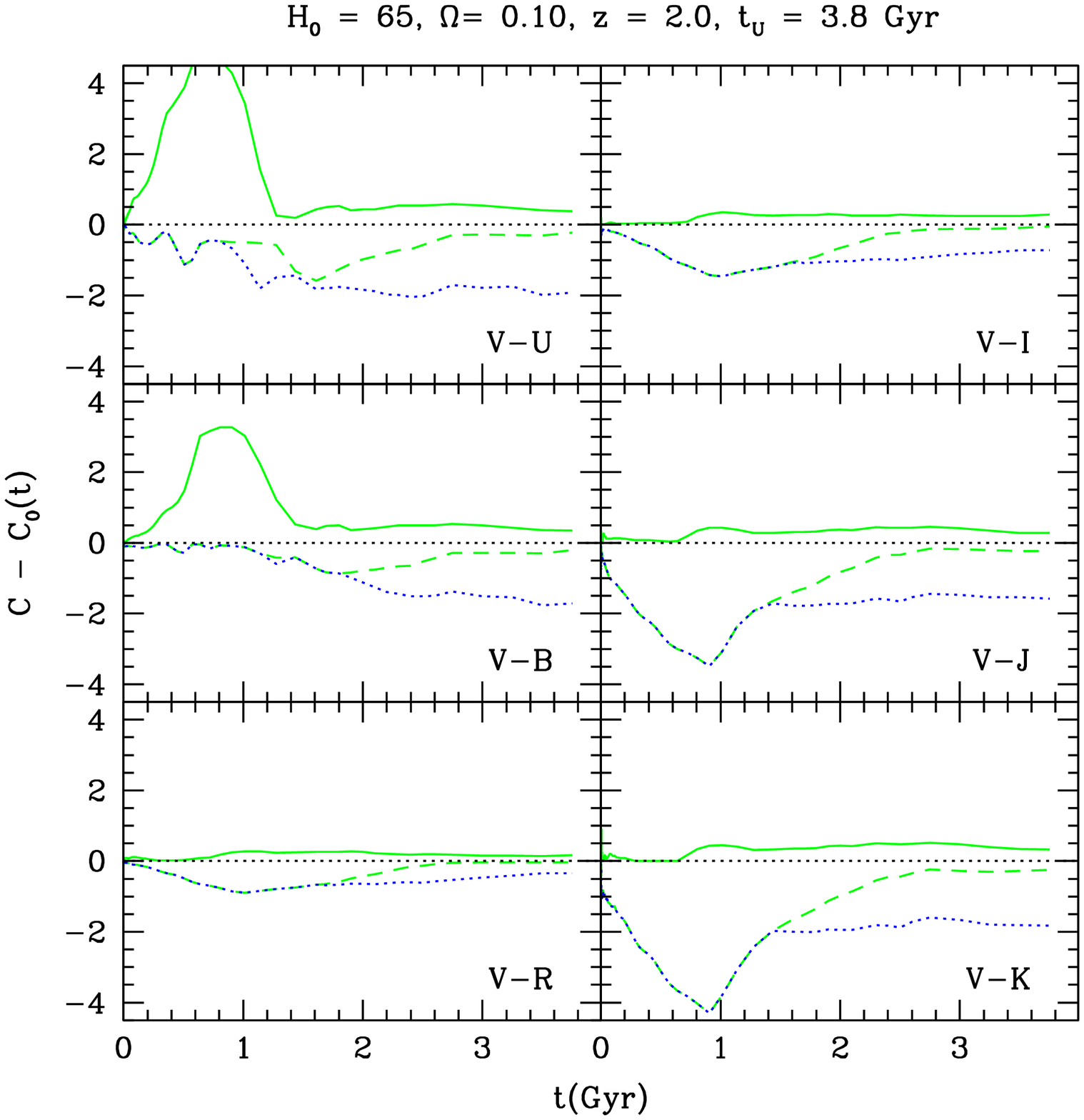}{10cm}{0}{50}{50}{-150}{-80}
\caption{
Color vs. time in the observer frame at $z=2$.
The color of the $Z=Z_\odot$ SSP model computed with the $P$ tracks, the
Salpeter IMF, and the Pickles stellar library have been subtracted from
each line. See \S11.6 for details.
}
\plotfiddle{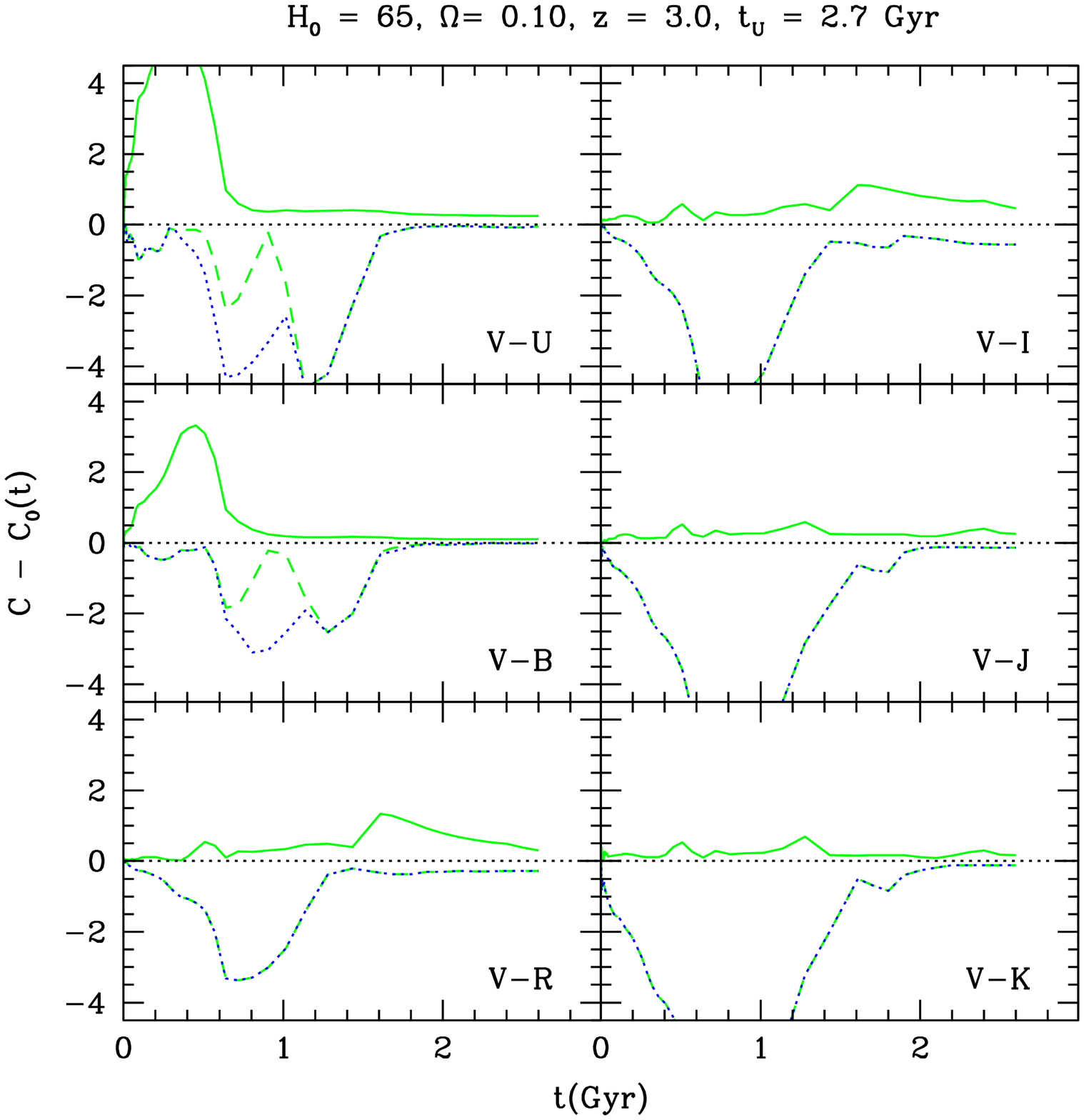}{10cm}{0}{50}{50}{-150}{-80}
\caption{
Color vs. time in the observer frame at $z=3$.
The color of the $Z=Z_\odot$ SSP model computed with the $P$ tracks, the
Salpeter IMF, and the Pickles stellar library have been subtracted from
each line. See \S11.6 for details.
}
\end{figure}

\begin{figure}
\plotfiddle{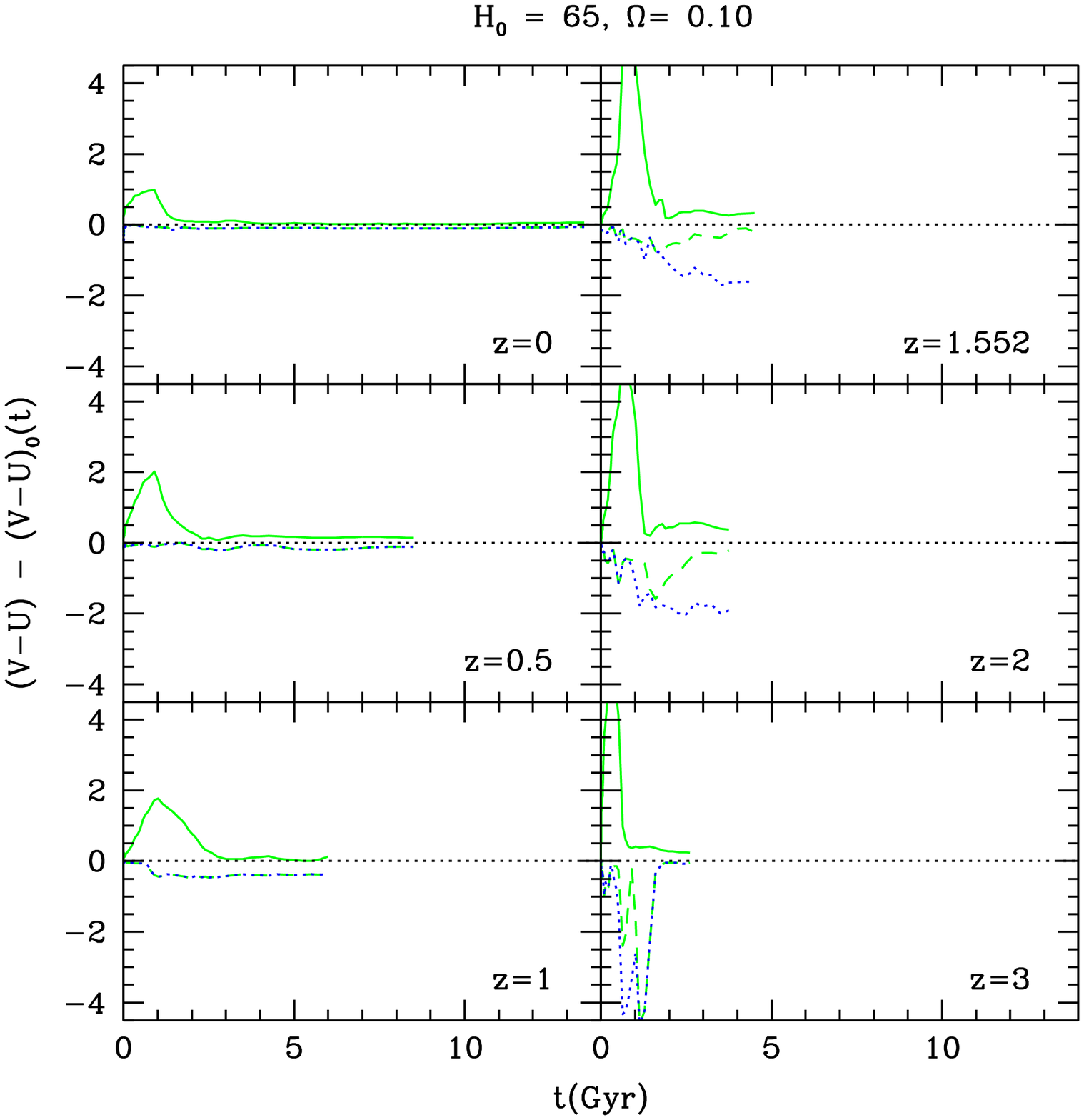}{10cm}{0}{50}{50}{-150}{-80}
\caption{
$(V-U)$ vs. time in the observer frame for various values of $z$.
The color of the $Z=Z_\odot$ SSP model computed with the $P$ tracks, the
Salpeter IMF, and the Pickles stellar library have been subtracted from
each line. See \S11.6 for details.
}
\plotfiddle{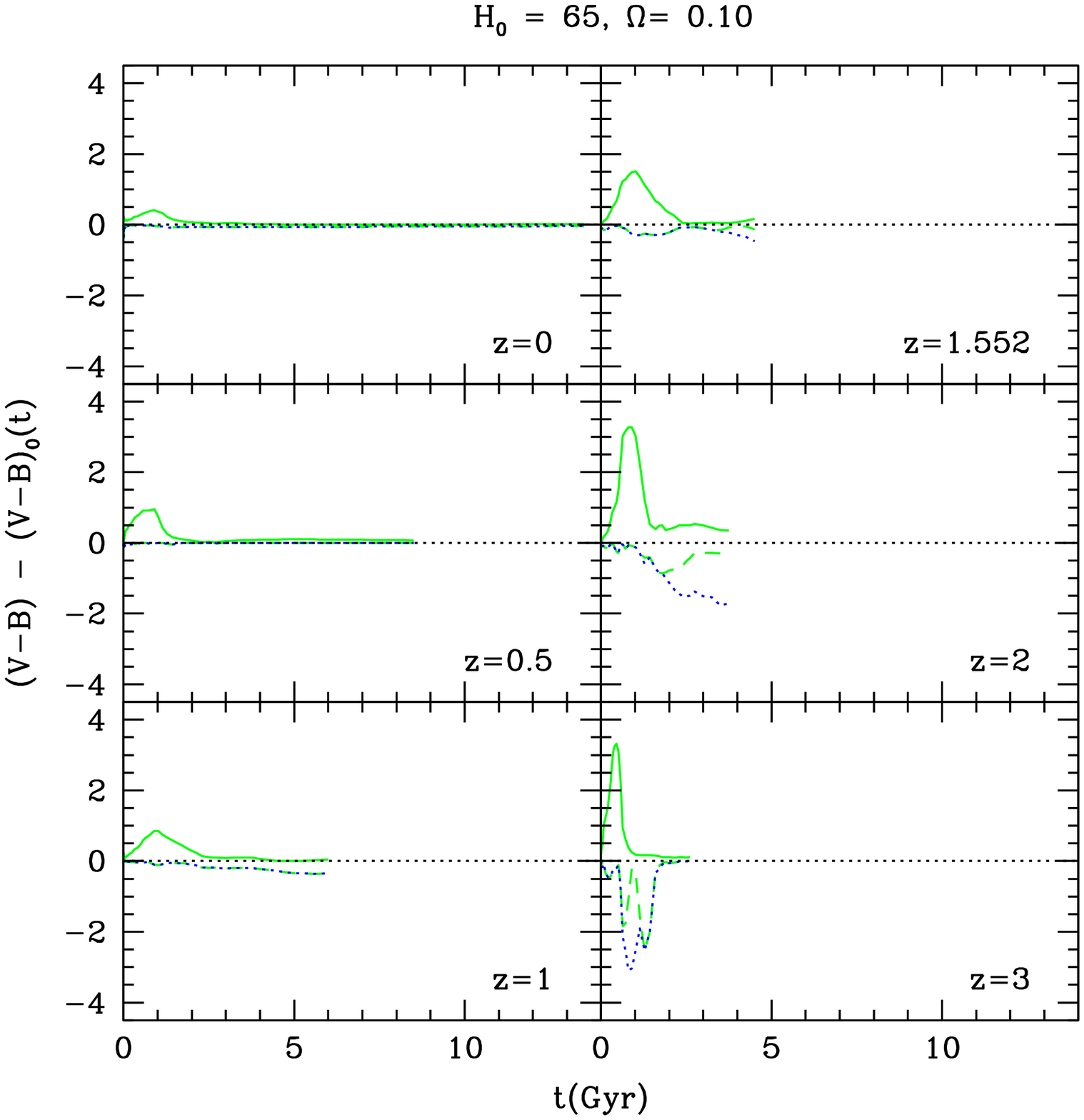}{10cm}{0}{50}{50}{-150}{-80}
\caption{
$(V-B)$ vs. time in the observer frame for various values of $z$.
The color of the $Z=Z_\odot$ SSP model computed with the $P$ tracks, the
Salpeter IMF, and the Pickles stellar library have been subtracted from
each line. See \S11.6 for details.
}
\end{figure}

\begin{figure}
\plotfiddle{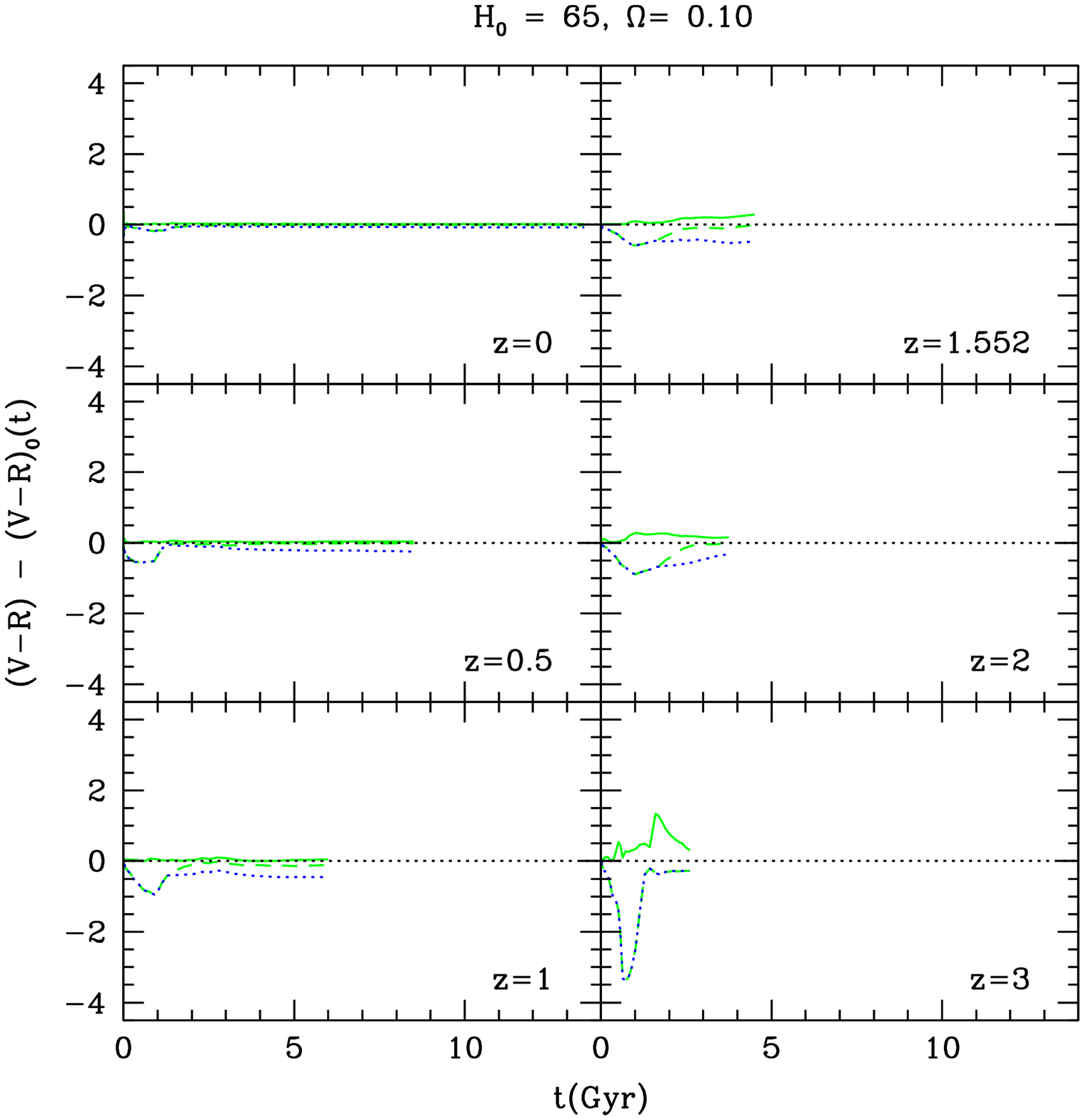}{10cm}{0}{50}{50}{-150}{-80}
\caption{
$(V-R)$ vs. time in the observer frame for various values of $z$.
The color of the $Z=Z_\odot$ SSP model computed with the $P$ tracks, the
Salpeter IMF, and the Pickles stellar library have been subtracted from
each line. See \S11.6 for details.
}
\plotfiddle{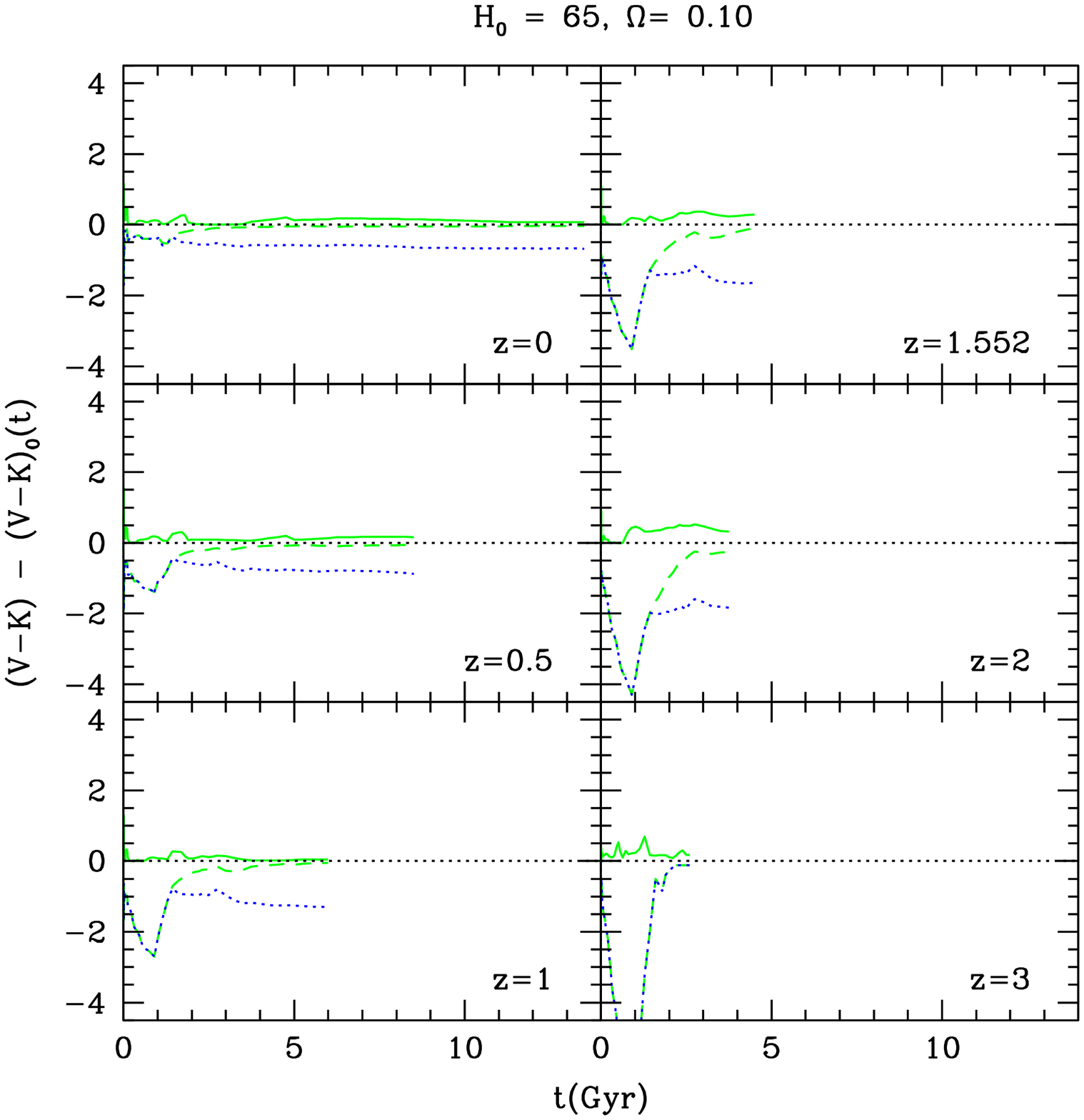}{10cm}{0}{50}{50}{-150}{-80}
\caption{
$(V-K)$ vs. time in the observer frame for various values of $z$.
The color of the $Z=Z_\odot$ SSP model computed with the $P$ tracks, the
Salpeter IMF, and the Pickles stellar library have been subtracted from
each line. See \S11.6 for details.
}
\end{figure}

\section{Summary and Conclusions}

Present population synthesis models show reasonable agreement
with the observed spectrum of stellar populations of various ages and
metal content. Differences in results from different codes can be understood
in terms of the different ingredients used to build the models and do
not necessarily represent violations of physical principles by some
of these models.
However, inspection of Fig. 20 shows that two different
sets of evolutionary tracks
for stars of the same metallicity produce models that at early ages differ
in brightness and color from 0.5 to 1 mag, depending on the specific bands.
The differences decrease at present ages in the rest frame, but are large
in the observer frame at $z > 2$. Thus any attempt to date distant galaxies,
for instance, based on fitting observed colors to these lines will produce
ages that depend critically on the set of models which is used. Note that
from $z$ of 3 to 3.5 $(V-K)$ in the two models differs by more than 1 mag.
This difference is produced by the corresponding difference between the
models seen in the rest frame at 10 Myr. From Figs. 15 and 16
these differences
can be understood in terms of the different contribution of the same
stellar groups to the total $V$ and $K$ flux in the two models.

Even though at the present age models built with different IMFs show reasonably
similar colors and brightness, the early evolution of these models is
quite different at early ages (Fig. 21), resulting in larger color
differences in the observer frame at $z > 2$. Thus, the more we know about
the IMF, the better the model predictions can be constrained.
The small color differences seen in the rest frame when different stellar
libraries of the same metallicity are used, are magnified in the observer frame
(Fig. 22).
When the $k$ correction brings opposing flux differences into each
filter, the difference in the resulting color is enhanced.
Fig. 23 shows the danger of interpreting data for one stellar system with
models of the wrong metallicity. The color differences between these models,
especially in the observer frame,
are so large as to make any conclusion thus derived very uncertain.

It is common practice to use solar metallicity models when no information
is available about the chemical abundance of a given stellar system.
Galaxies evolving according the $Z(t)$ laws of Fig. 24 show
color differences with respect to the $Z=Z_\odot$ model which are not
larger than the differences introduced by the other sources of uncertainties
discussed so far. Hence, the solar metallicity approximation may be
justified in some instances. The color differences between the chemically
inhomogeneous composite population and the purely solar case (Fig. 25),
are much smaller than the ones shown in Fig. 23
for chemically homogeneous SSPs.

Figs. 26 to 39 indicate that some colors, especially $(V-R)$, when
measured in the observer frame are less sensitive to model predictions
than other colors.
From Figs. 26 to 31, metallicity $Z$ and the SFR are the
most dominant factors determining the range of allowed colors.

I expect that through these simple examples the reader can get a feeling
of the kind of uncertainties introduced by the many ingredients entering
the stellar population synthesis problem, and that he or she will be
motivated to try his or her own error estimates when using these models.

{}


\begin{thebibliography}{}

\bibitem[]{}
Allard, F., \& Hauschildt, P.H. 1995, ApJ, 445, 433
\bibitem[]{}
Alongi, M., Bertelli, G., Bressan, A., Chiosi, C., Fagotto, F., Greggio, L., \& Nasi, E. 1993, A\&AS, 97, 851
\bibitem[]{}
Arag\'on-Salamanca, A., Ellis, R.S.E., Couch, W. J., Carter, D. 1993, MNRAS, 262, 764A
\bibitem[]{}
Arimoto, N., \& Yoshii, Y. 1987, A\&A, 173, 23
\bibitem[]{}
Barbuy, B., Ortolani, S., Bica, E., Renzini, A., \& Guarnieri, M.D. 1997, in IAU Symp. 189, {\it Fundamental Stellar Parameters: Confrontation Between Observation and Theory}, eds. J. Davis, A. Booth \& T. Bedding, Kluwer Acad. Pub., p. 203
\bibitem[]{}
Bender, R., Ziegler, B., \& Bruzual A., G. 1996, ApJ Letters, 463, L51
\bibitem[]{}
Bessell, M.S., Brett, J., Scholtz, M., \& Wood, P. 1989, A\&AS, 77, 1
\bibitem[]{}
---------. 1991, A\&AS, 89, 335
\bibitem[]{}
Bica, E., \& Alloin, D. 1986, A\&A, 162, 21
\bibitem[]{}
---------. 1987, A\&A, 186, 49
\bibitem[]{}
Bica, E., Alloin, D., \& Schmitt, H. 1994, A\&A, 283, 805
\bibitem[]{}
Bica, E., Alloin, D., Bonatto, C., Pastoriza, M.G., Jablonka, P., Schmidt, A., \& Schmitt, H.R. 1996a, in {\it A Data Base for Galaxy Evolution Modeling}, eds. C. Leitherer et al., PASP, 108, 996
\bibitem[]{}
Bica, E., Clari\'a, J.J., Dottori, H., Santos Jr., J.F.C., Piatti, A. E. 1996b, ApJS, 102, 57
\bibitem[]{}
Bressan, A., Chiosi, C., \& Fagotto, F. 1994, ApJS, 94, 63
\bibitem[]{}
Bressan, A., Fagotto, F., Bertelli, G., \& Chiosi, C. 1993, A\&AS, 100, 647
\bibitem[]{}
Bruzual A., G. 1998, in {\it The Evolution of Galaxies on Cosmological Time scales}, eds. J.E. Beckman and T.J. Mahoney, ASP Conference Series, Vol. 187, p. 245
\bibitem[]{}
---------. 1999, in {\it The Hy-Redshift Universe: Galaxy Formation and Evolution at High Redshift}, eds. A. J. Bunker and W. J. M. van Breugel, ASP Conference Series, Vol. 193, p. 121
\bibitem[]{}
---------. 2000, in {\it Euroconference on The Evolution of Galaxies, I- Observational Clues}, eds. J.M. V{\'\i}lchez, G. Stasinska, and E. P\'erez, Kluwer Academic Publisher, in press
\bibitem[]{}
Bruzual A., G., Barbuy, B., Ortolani, S., Bica, E., Cuisinier, F., Lejeune, T., \& Schiavon, R. 1997, AJ, 114, 1531
\bibitem[]{}
Bruzual A., G. \& Charlot, S. 1993, ApJ, 405, 538 (BC93)
\bibitem[]{}
---------. 2000, ApJ, in preparation (BC2000)
\bibitem[]{}
Burstein, D., Bertola, F., Buson, L.M., Faber, S.M., and Lauer, T.R. 1988, ApJ, 328, 440
\bibitem[]{}
Buzzoni, A. 1989, ApJS, 71, 817
\bibitem[]{}
---------. 1999, in IAU Symposium No. 183 {\it Cosmological Parameters and the Evolution of the Universe}, ed. K. Sato, Dordrecht: Kluwer, p. 134
\bibitem[]{}
Charlot, S., and Bruzual A., G. 1991, ApJ, 367, 126 (CB91)
\bibitem[]{}
Cool, A.M. 1997, in {\it Advances in Stellar Evolution}, eds. R. T. Rood and A. Renzini, Cambridge University Press, p. 191
\bibitem[]{}
D'Antona, F. 1999, in {\it The Galactic Halo: from Globular Clusters to Field Stars}, 35th Liege Int. Astroph. Colloquium, astro-ph/9910312
\bibitem[]{}
Eggen, O.J., and Sandage, A.R. 1964, ApJ, 140, 130
\bibitem[]{}
Fagotto, F., Bressan, A., Bertelli, G., \& Chiosi, C. 1994a, A\&AS, 100, 647
\bibitem[]{}
---------. 1994b, A\&AS, 104, 365
\bibitem[]{}
---------. 1994c, A\&AS, 105, 29
\bibitem[]{}
Fluks, M. et al. 1994, A\&AS, 105, 311
\bibitem[]{}
Fritze-v.Alvensleben, U. \& Gerhard, O.E. 1994, A\&A, 285, 751
\bibitem[]{}
Gilliland, R.L., Brown, T.M., Duncan, D.K., Suntzeff, N.B., Wesley Lockwood, G., Thompson, D.T., Schild, R.E., Jeffrey, W.A., and Penprase, B.E., 1991, AJ, 101, 541
\bibitem[]{}
Girardi, L., Bressan, A., Chiosi, C., Bertelli, G., \& Nasi, E. 1996, A\&AS, 117, 113
\bibitem[]{}
Girardi, L., Bressan, A., Bertelli, G., \& Chiosi, C. 2000, A\&AS, 141, 371
\bibitem[]{}
Greggio, L., and Renzini, A., 1990, ApJ, 364, 35
\bibitem[]{}
Guarnieri, M.D., Ortolani, S., Montegriffo, P., Renzini, A.,
\bibitem[]{}
Guiderdoni, B. \& Rocca-Volmerange, B. 1987, A\&A, 186, 1
\bibitem[]{}
Gunn, J.E., and Stryker, L.L., 1983, ApJS, 52, 121
\bibitem[]{}
Iglesias, C.A., Rogers, F.J., \& Wilson, B.G. 1992, ApJ, 397, 717
\bibitem[]{}
Janes, K.A., 1985, in Calibration of Fundamental Stellar Quantities, IAU Symposium No. 111, D.S. Hayes, L.E. Pasinetti, and A.G. Davis Philip, (Dordrecht: Reidel), 361
\bibitem[]{}
Janes, K.A., and Smith, G.H., 1984, AJ, 89, 487
\bibitem[]{}
Kaluzny, J. 1997, A\&AS, 121, 455
\bibitem[]{}
King, I.R., Anderson, J., Cool, A.M., Piotto, G. 1998, ApJ, 492, L37
\bibitem[]{}
Kroupa, P., Tout, C.A., \& Gilmore, G. 1993, MNRAS, 262, 545
\bibitem[]{}
Kurucz, R. 1995, private communication
\bibitem[]{}
Lejeune, T., Cuisinier, F., \& Buser, R. 1997, A\&AS, 125, 229 (LCB97)
\bibitem[]{}
---------. 1998, A\&AS, 130, 65 (LCB98)
\bibitem[]{}
Metcalfe, N., Shanks, T., Fong, R., Gardner, J., Roche, N. 1996, IAU Symp. 171, p. 225
\bibitem[]{}
Micela, G., Sciortino, S., Vaiana, G.S., Schmitt, J.H.M.M., Stern, R.A., Harnden, F.R., Jr., and Rosner, R., 1988, ApJ, 325, 798
\bibitem[]{}
Miller, G.E. \& Scalo, J.M. 1979, 41, 513
\bibitem[]{}
Peterson, D.M., \& Solensky, R. 1988, ApJ, 333, 256
\bibitem[]{}
Pickles, A.J. 1998, PASP, 110, 863
\bibitem[]{}
Pozzetti, L., Bruzual A., G., Zamorani, G. 1996, MNRAS, 281, 953
\bibitem[]{}
Racine, R., 1971, ApJ, 168, 393
\bibitem[]{}
Renzini, A., 1981, Ann. Phys. Fr., 6, 87
\bibitem[]{}
Salpeter, E.E. 1955, ApJ, 121, 161
\bibitem[]{}
Santos, J.F.C.Jr., Bica, E., Dottori, H., Ortolani, S., \& Barbuy, B. 1995, A\&A, 303, 753
\bibitem[]{}
Scalo, J.M. 1986, Fund. Cosmic Phys, 11, 1
\bibitem[]{}
Spinrad, H., Dey, A., Stern, D., Dunlop, J., Peacock, J., Jim\'enez, R., Windhorst, R. 1997, ApJ, 484, 581
\bibitem[]{}
Stanford, S.A., Eisenhardt, P.R., \& Dickinson, M. 1995, ApJ, 450, 512
\bibitem[]{}
---------. 1998, ApJ, 492, 461
\bibitem[]{}
Upgren, A.R., 1974, ApJ, 193, 359
\bibitem[]{}
Upgren, A.R., and Weis, E.W., 1977, AJ, 82, 978
\bibitem[]{}
Worthey, G. 1994, ApJS, 95, 107
\bibitem[]{}
Worthey, G., Faber, S.M., Gonz\'alez, J.J., \& Burstein, D. 1994, ApJS, 94, 687

\end{thebibliography}
\end{document}